%% file: main.tex
\documentclass[letterpaper,twocolumn,10pt]{article}
\usepackage{usenix2019_v3}

\usepackage{tikz}

\usepackage{amsmath}

\usepackage{filecontents}
\usepackage{xurl}
\usepackage{float}
\usepackage[]{graphicx}

\usepackage[symbol]{footmisc}

\usepackage{booktabs}

\usepackage{subfig}

\definecolor{dark green}{RGB}{94,145,40}
\definecolor{dark yellow}{RGB}{210,200,40}

\newcommand{\reviewed}[1]{\textcolor{black}{#1}}

\providecommand{\ie}{\emph{i.e.,} }
\providecommand{\eg}{\emph{e.g.,} }

\begin{document}

\date{}

\newcommand\ChapterPrecis[2]{%
\begin{tikzpicture}[remember picture,overlay]
\node[anchor=north, draw=black, fill=yellow!20, inner sep=5pt, rounded corners, yshift=-#1] at (current page.north) 
{\parbox[t][1.3cm][c]{\textwidth}{\small #2}};
\end{tikzpicture}%
}

\title{\Large \bf The Conspiracy Money Machine: Uncovering\\ Telegram's Conspiracy Channels and their Profit Model
}
\author{
{\rm Vincenzo Imperati}\\
\text{imperati@di.uniroma1.it}\\
Sapienza University of Rome 
\and
{\rm Massimo La Morgia}\\
\text{lamorgia@di.uniroma1.it}\\
Sapienza University of Rome
\and 
{\rm Alessandro Mei}\\
\text{mei@di.uniroma1.it}\\
Sapienza University of Rome
\and
{\rm Alberto Maria Mongardini}\\
\text{mongardini@di.uniroma1.it}\\
Sapienza University of Rome
\and 
{\rm Francesco Sassi}\\
\text{sassi@di.uniroma1.it}\\
Sapienza University of Rome
} 


\maketitle

\ChapterPrecis{0.8cm}{If you cite this paper, please use the USENIX Security reference: Vincenzo Imperati, Massimo La Morgia, Alessandro Mei,
Alberto Maria Mongardini, and Francesco Sassi. 2025. "The Conspiracy Money Machine: Uncovering Telegram's Conspiracy Channels and their Profit Model." In 34th USENIX Security Symposium (USENIX Security 25), pp. 5229-5246. 2025. \newline \url{https://www.usenix.org/system/files/usenixsecurity25-imperati.pdf}
}

\begin{abstract}

In recent years, major social media platforms have implemented increasingly strict moderation policies, resulting in bans and restrictions on conspiracy theory-related content. To circumvent these restrictions, conspiracy theorists are turning to alternatives, such as Telegram, where they can express and spread their views with fewer limitations. Telegram offers channels—virtual rooms where only administrators can broadcast messages---and a more permissive content policy. These features have created the perfect breeding ground for a complex ecosystem of conspiracy channels.

In this paper, we illuminate this ecosystem. First, we propose an approach to detect conspiracy channels. Then, we discover that conspiracy channels can be clustered into four distinct communities comprising over 17,000 channels.
Finally, we uncover the "Conspiracy Money Machine," revealing how most conspiracy channels actively seek to profit from their subscribers. We find conspiracy theorists leverage e-commerce platforms to sell questionable products or lucratively promote them through affiliate links. Moreover, we observe that conspiracy channels use donation and crowdfunding platforms to raise funds for their campaigns. We determine that this business involves hundreds of thousands of donors and generates a turnover of almost \$71 million. 
\end{abstract}

\section{Introduction}
Conspiracy theories have been an integral part of human history, offering alternative interpretations for complex events~\cite{van2017conspiracy}. The most common definition, which we use in our work,  is that a conspiracy theory is a belief that an event or situation is the result of a secret plan orchestrated by powerful people or organizations~\cite{ballard2022conspiracy,phadke2021makes}. 
A notorious example is the Flat Earth theory~\cite{flatearth}. Despite centuries of scientific evidence proving the Earth's roundness, the theory continues to be discussed and promoted by several communities~\cite{mohammed2019conspiracy,fernbach2023conspiracy}.
Throughout history, several conspiracy theories have emerged on a wide range of topics, like the Moon Landing Hoax~\cite{moonlandinghoax}, JFK Assassination~\cite{willman1998traversing}, Holocaust Denial~\cite{van2017conspiracy}, Elvis Presley's Faked Death~\cite{clarke2019conspiracy}, and 9/11 Conspiracy Theories~\cite{stempel2007media}. 

Nowadays, with the advent of the Internet and social media, conspiracy theories have found new outlets to spread and gain traction~\cite{cinelli2022conspiracy}.
A notable example is the Pizzagate conspiracy theory, which originated and spread on online bulletin boards in 2016~\cite{tuters2018post}.
In 2017, online forums acted as a catalyst for QAnon conspiracy theories~\cite{de2020tracing}, which alleged that a global cabal of malevolent elites was involved in heinous activities. 
The advent of the COVID-19 pandemic has sparked various online conspiracy theories, including claims that the virus is a bio-weapon for population control~\cite{imhoff2020bioweapon}, and that 5G technology is somehow linked to the spread of the virus~\cite{ahmed2020covid}.
Finally, on January 6, 2021, a pro-Trump mob stormed the U.S. Capitol building, disrupting the certification of the 2020 presidential election results~\cite{muhammad2021study}.
These incidents led the major social media platforms to implement content moderation to curb the dissemination of these theories~\cite{twitterbanqanon}. In response, conspiracy theorists are flocking to less moderated platforms to freely share their views.
Anecdotal evidence from various news sources~\cite{telegramconspiracy,telegramconspiracy1,telegramconspiracy2} underscores Telegram, one of the most popular instant messaging applications, as one such platform.
This is not surprising, as Telegram offers a permissive content policy and channels---virtual rooms where the admins can broadcast messages to large audiences.

In our work, we perform a large-scale study of Telegram to shed light on its ecosystem of conspiracy channels.
We propose a novel approach to identify channels related to conspiracy theories by examining the URLs they share.
In particular, we leverage previous scientific work on conspiracy theories to build the Conspiracy Resource Dataset, which contains a list of online resources (\eg YouTube videos, Reddit posts) linked to conspiracy theories. Then, we use the TGDataset, a public dataset of over 120,000 Telegram channels, to find channels sharing conspiracy-related URLs with their subscribers.
Then, we utilize a community detection algorithm to analyze Telegram communities, finding that conspiracy-related channels are clustered in four specific communities. We characterize these communities by analyzing their language and most influential channels. We refer to the channels contained in these communities as the Conspiracy Channel Dataset.
The analysis of the Conspiracy Channel Dataset highlighted the presence of channels actively seeking to profit from their subscriber.
We characterize and quantify this phenomenon, focusing on three monetization strategies: donations, crowdfunding campaigns, and affiliate programs.
First, we focus on analyzing donation and crowdfunding platforms. While we could not extract information about donation URLs, we find several insights about crowdfunding campaigns. Indeed, crowdfunding projects sponsored by conspiracy channels collected millions of dollars donated by over 900,000 backers. Moreover, analyzing the top-funded campaigns, we find they are linked to far-right support, COVID-19 restriction opposition, and truth-revealing documentaries against governments and powerful individuals. Finally, we also find fake charity campaigns that are outright scams. 
Then, we find that conspiracy theorists exploit the lenient product policies of eBay, Teespring, and Etsy to promote questionable items to their subscribers, such as 5G shields, EMF stone protectors, and healing wands. Moreover, they exploit Amazon's tolerant book content policies to self-publish and profit from books claiming to ``reveal the truth'' about several topics.
Our work makes the following contribution:
\begin{itemize}
    \item \textbf{Conspiracy Datasets.}
     We release two datasets~\cite{anonymous_groundtruth}. The first one is the Conspiracy Resource Dataset, a collection of conspiracy-related web resources gathered through an extensive literature review. The second is the Conspiracy URLs Dataset, a list of 177,567 resolved unique URLs shared by conspiracy theory-related channels. We believe these datasets can enable further studies on identifying and characterizing conspiracy communities on other platforms and determining their activities.
    \item \textbf{Conspiracy Detection and Analysis.}
    We propose an approach to identify conspiracy communities on Telegram, finding four large communities comprising 17,829 channels. We characterize each community by analyzing their language and their most influential channels.

    \item \textbf{Conspiracy Monetization.} 
    Finally, we identify the potential strategies that conspiracy theorists can employ to generate revenue from the subscribers of Telegram channels.
    We find that the most popular approaches involve donations, crowdfunding campaigns, and affiliate programs. 
    We discover 123K URLs linked to donation platforms and more than 24K URLs related to crowdfunding campaigns. Quantifying the amount of money raised with projects sponsored by conspiracy channels, we discovered that they amassed almost \$71M donated by over 906K backers.

\end{itemize}

\section{How Telegram works}
Telegram is one of the most prominent instant messaging application platforms, with over 700 million active users in 2023~\cite{TelegramUsers}. 
Like most instant messaging applications, Telegram provides one-to-one messaging. Indeed, Telegram users can easily engage in conversations by exchanging text messages, multimedia content, and files.
Moreover, users can also create and join groups---chats where any member can post content. This feature allows users to create communities around shared interests for discussions, event planning, and coordination. 

\textbf{Channels.} One of the core features of Telegram is channels, chats designed to provide one-to-many messaging. Indeed, the only user who can send messages into a channel is the admin. Other Telegram users can freely join a channel and read its posts but cannot send messages.
This feature allows the admin to share content with a huge number of subscribers, making Telegram channels a prime broadcasting medium for disseminating news and announcements. Channels on Telegram are identified by unique usernames, have a title, and may include a description and a chat picture. Moreover, while group members can see which users are in their group, only a channel admin can access the list of subscribers.

\textbf{Message forwarding} Another core functionality for distributing content within Telegram is message forwarding. Indeed, users (or admins of a channel) can easily forward a message from one chat to another. The forwarded message displays the original message's author, serving as a bridge between groups, channels, and private chats.

\section{Related work}
\label{sec:sota}

Telegram has recently gained substantial attention, with several prior studies investigating questionable activities on the platform Weerasinghe et al.\cite{weerasinghe2020pod} studied ``pods'', organized groups created to artificially boost Instagram popularity. Other studies have explored Telegram's use in cryptocurrency market manipulations like pump and dump\cite{la2023doge} and Ponzi schemes~\cite{nizzoli2020charting}. Additionally, researchers have noted its misuse by terrorist organizations for propaganda and recruitment~\cite{yayla2017telegram,shehabat2017encrypted}. Research on conspiracy theories on Telegram is limited, with most studies focusing on other social media platforms. Here is a report on these studies.

\textbf{Telegram.} 
Hoseini et al.~\cite{hoseini2023globalization} examined 161 QAnon groups on Telegram, analyzing their toxicity and performing topic modeling to understand the QAnon narrative in multiple languages.
Garry et al.~\cite{garry2021qanon} focuses on analyzing 35 QAnon Telegram channels,  discovering that they spread disinformation messages to recruit new adepts. La Morgia et al.~\cite{la2018pretending,lamorgia2023trap} analyze over 120,000 Telegram channels, focusing on detecting fakes and clones. They discovered that these channels are used to lure users into conspiracy-related channels.
Unlike these studies, we examine the overall landscape of conspiracy-related channels on Telegram, propose a method to identify their communities, and analyze their profit model.

\textbf{YouTube.} Leidwich et al.~\cite{ledwich2020algorithmic} explore whether YouTube's recommendation algorithm promotes radicalization by guiding users to increasingly extreme content. They categorized 816 channels, including 79 conspiracy ones.  
Clark et al.~\cite{clarkunderstanding} leverage the dataset of~\cite{ledwich2020algorithmic} to find YouTube communities. They create an embedding for the channels considering the channel's subscribers. Then, they leverage cluster algorithms to reveal the  YouTube communities, finding QAnon and conspiracy-related ones.

\textbf{Reddit.} Phadke et al.\cite{phadke2022pathways} analyzed conspiracy theories on Reddit using the RECRO model to study user radicalization. They examined 169 million contributions from 36,000 users, identifying four engagement trends. Papasavva et al.\cite{aliapoulios2021gospel} focused on the QAnon conspiracy, analyzing 4,949 "Q drops" and their dissemination on Reddit, finding continued sharing even after QAnon subreddits were banned.
Engel et al.~\cite{engel2022characterizing} analyze the submission of 13K users in  19 QAnon-related subreddits. They discover they are active across various subreddits, often posting harmful content from low-quality sources.
Phadke et al.~\cite{phadke2021makes} analyze 56 conspiracy communities on Reddit, creating a ground truth of 60k users to develop a machine learning model to predict if a Reddit user will eventually join conspiracy communities.

\textbf{Voat.} Voat, a Reddit clone, gained notoriety after Reddit banned Pizzagate and QAnon-related subreddits~\cite{qanonvoat}. Papasavva et al.\cite{papasavva2021qoincidence} analyzed over 150,000 posts from the largest QAnon forum on Voat, finding a focus on Trump and US politics. Mekacher et al.\cite{mekacher2022can} created a dataset of over 2.3 million Voat submissions, discovering that many active subverses centered on hate speech and conspiracy theories.

\textbf{4chan/8kun.} 
4chan~\cite{bernstein20114chan}, an image-based bulletin board, has been linked to conspiracy theories, notably the PizzaGate conspiracy~\cite{bleakley2023panic}. Papasavva et al.\cite{papasavva2020raiders} analyzed over 3.5 million messages on 4chan's \textit{/pol} board, finding antisemitic conspiracy theories. Similar content is found on 8kun, a platform associated with white supremacism and hate crimes\cite{hanley2022no}. Papasavva et al.~\cite{aliapoulios2021gospel} discovered that 8kun QAnon threads are significantly larger than those on 4chan.

\reviewed{
\textbf{Monetization misuse and frauds.}  Ballard et al.~\cite{ballard2022conspiracy} leverage the datasets from~\cite{ledwich2020algorithmic,clarkunderstanding} to investigate the monetization strategies of YouTube conspiracy channels. They find that these channels have a high prevalence of predatory or deceptive ads, are often demonetized, and use alternative income sources via third-party platforms. 
Broniatowski et al.\cite{broniatowski2023measuring} conducted a study involving 1,448 respondents and found that most are unlikely to pay for online conspiracy content. However, respondents who avoid mainstream media and rely on social media for news are more inclined to pay for such content.
Chachra et al.~\cite{chachra2015affiliate}  study cookie-stuffing,  a technique used to divert revenue commissions in affiliate marketing networks. They highlight the fraud mechanisms and identify which categories of merchants are most targeted, noting that scammers employ a wide range of evasive techniques.
}
\section{Methodology}
To uncover the structure of channels in Telegram that are related to conspiracy theories, we build a methodology consisting of three steps: First, data collection (including the introduction of a new dataset); second, detection of "conspiracy channels;" third, identification of communities of channels linked to conspiracy theories.

\subsection{Data collection}
We leverage two datasets. The first one is the TGDataset~\cite{la2023tgdataset}, the largest collection of public Telegram channels, with over 120,000 channels and 400 million messages. \reviewed{The dataset contains all the messages shared by the collected channels until July 31, 2022.} This dataset provides information about the channels (\eg their title, description, and creation date), all the messages sent with their timestamp and whether a message has been forwarded, and from which channel it originated.
Then, we build a novel dataset, the Conspiracy Resource Dataset, that we describe in the following.

\subsubsection{Conspiracy Resource Dataset}
This dataset is a collection of conspiracy-related resources extracted from an extensive review of the previous works about conspiracy theories reported in Section~\ref{sec:sota}.
To construct this dataset, we focus on studies that provide explicit pointers to the sources they analyze, either within the manuscript or in dedicated repositories. In the following, we report for each platform the number of resources we find and the reference article:
\begin{itemize}
    \item \textbf{YouTube.}
    We follow the approach of~\cite{ballard2022conspiracy} and use two repositories of YouTube channels reported in~\cite{ledwich2020algorithmic, clarkunderstanding}. These repositories contain a list of 4,007 YouTube channels manually labeled as conspiracy-related.
    For each channel, we extract the complete list of their videos, resulting in a total of 1,973,428 video IDs.
    \item \textbf{Reddit.}
    We leverage the work of~\cite{aliapoulios2021gospel,phadke2022pathways,engel2022characterizing,phadke2021makes} to collect a list of 91 subreddits identified as conspiracy-related.
    \item \textbf{Voat.} We consider 3 Voat subverses related to the QAnon conspiracy theory reported in~\cite{papasavva2021qoincidence,mekacher2022can}. 
    \item \textbf{4chan/8kun.}
    We did not find any resource specifically related to conspiracy theories on 4chan from previous work. Indeed, the infamous \textit{/pol} discussion board is too general and discusses as well topics unrelated to conspiracy theories. Instead, we collect a list of 7 boards related to QAnon on 8kun from the work in~\cite{aliapoulios2021gospel}.
    \item \textbf{Websites.} 
    From the work of Papasavva et al. ~\cite{aliapoulios2021gospel}, we collect 8 websites that are well-known aggregators of Q drops spread by the QAnon conspiracy. Then, we use OpenSources~\cite{FakeNewsCorpus} (also used to study QAnon in~\cite{hanley2022no}) to extract 120 website domains linked to conspiracy theories.
\end{itemize}

Although there are other works mentioning conspiracy theories in other platforms like 
Parler~\cite{aliapoulios2021large,bar2023finding} or Twitter~\cite{ahmed2020covid}, they do not publicly release their datasets or provide URLs related to conspiracy theories that we can use in our study.
Tab~\ref{tab:conspiracy_groundtruth} reports all the resources we find and the related papers.
Moreover, we release in~\cite{anonymous_groundtruth} the full dataset of collected resources.

\begin{table}
   \centering
   \small
   \caption{%
       Summary of the conspiracy-related resources we find in previous work and related URLs we extract from Telegram. 
  }\label{tab:conspiracy_groundtruth}
   \begin{tabular}{l l l r }
      \toprule
       Paper & Type & \# Resources & \#  URLs \\
      \midrule
      \cite{ledwich2020algorithmic,ballard2022conspiracy, clarkunderstanding} & YouTube & 4,007 channels & 146,915\\ 
      \cite{aliapoulios2021gospel,phadke2022pathways,engel2022characterizing,phadke2021makes} & Reddit & 91 subreddits & 17,870 \\ 
      \cite{papasavva2021qoincidence,mekacher2022can} & Voat & 3 subverses & 21 \\ 
      \cite{aliapoulios2021gospel} & 8kun & 7 boards & 3,768 \\ 
      \cite{aliapoulios2021gospel,hanley2022no} & Web  & 128 websites & 297,170 \\ 
    
      \bottomrule
    \end{tabular}
\end{table}

\subsection{Conspiracy channels detection}
\label{sec:detection-methodology}
We devise a methodology that combines the TGDataset and the Conspiracy Resource Dataset to find channels on Telegram related to conspiracy theories.
The detection is performed in four steps: First, we extract and pre-process URLs from the TGDataset, and we perform the match with the resources found in the Conspiracy Resource Dataset. Then, we use graph analysis to find clusters of conspiracy-related channels, and, finally, we validate our results.
\subsubsection{Data extraction and pre-processing}
We parse all the messages (498,320,597) in the 120,979 TGDataset channels and use regular expressions to extract all the URLs. In this way, we obtain 205,046,775 URLs,  84,809,578 of which are unique.
A first analysis of the URLs reveals that 20.2\% (17,140,343 URLs) have been shortened using URL shortener services such as \textit{bit.ly} (2,368,953 occurrences) or \textit{if.tt} (1,462,885 occurrences).
Since our methodology for detecting conspiracy theories channels revolves around identifying URLs associated with conspiracies, we want to ensure we do not miss any of them because it has been shortened.
\reviewed{Thus, between July and September 2023, we resolved the shortened URLs by sending HEAD requests to the shortening service using the Python Requests library.}

\subsubsection{URLs matching}
\label{sec:urls_matching}
Then, we extract URLs associated with the resources collected in the Conspiracy Resource Dataset.
We start by using the conspiracy channels and videos' IDs in the Conspiracy Resource Dataset to search YouTube URLs related to conspiracy theories. We detect 2,284 URLs (371 unique) linking to conspiracy channels and 144,631 URLs (52,120 unique) linking to conspiracy videos. 
The most widely shared channel is \textit{Fall Cabal}, 
a channel associated with the "Fall of the Cabal"~\cite{fallodthecabal}, an antisemitic documentary used to recruit QAnon followers affiliated with Dutch conspiracy theorist Janet Ossebaard~\cite{meder2021online}.
Instead, for Reddit, 8kun, and Voat, we extract all the URLs linking to conspiracy subreddits, boards, and subverses, respectively. We find 17,870 Reddit URLs (17,156 unique), 3,768 8kun URLs (2,509 unique), and 21 Voat URLs (11 unique).
Most of the URLs we find are related to the \textit{r/conspiracy} subreddit, the largest conspiracy theory discussion board on Reddit~\cite{phadke2022pathways}.

Finally, we extract all the URLs having the domain of the flagged websites, finding 297,170 URLs (105,877 unique) from 94 different resources.
The website providing the most matches is Zerohedge, a far-right news aggregator known for spreading conspiracy theories, particularly about COVID-19 ~\cite{zerohedgeban}. The second most popular is InfoWars, well-known for promoting conspiracy theories and fake news~\cite{zeng2021conceptualizing}. We also detect over 10k URLs linking to the \textit{qagg.news} website, a popular repository that stores the messages of the QAnon conspiracy theory~\cite{garry2021qanon,aliapoulios2021gospel}.

In total, we find 465,744 URLs (178,044 unique) posted by 11,487 Telegram channels. We publicly release the extracted URLs in~\cite{anonymous_groundtruth} to enable further studies. In the following, we will refer to this dataset as \textbf{Conspiracy URLs Dataset}.
Tab.~\ref{tab:conspiracy_groundtruth} succinctly reports the number of URLs collected inside the Telegram channels divided by resource type.

\subsubsection{Clustering conspiracy channels}
\label{sec:graph_analysis}
The previous analysis shows 11,487 Telegram channels that posted at least one message with a link to a conspiracy-related resource. 
We study how and if these channels are connected to better understand the phenomenon.
To do so, we follow the approach of~\cite{la2018pretending} and build the Telegram forwarding graph for the whole dataset of channels. 
The forwarding graph is a graph $G = (V,E) $ in which nodes in $G$ are channels and an edge $u \to v$ in $E$ represents the presence in $u$ of a message forwarded from channel $v$. Users of channel $u$ can follow the forwarded message and reach channel $v$. Thus, edges represent the possible flows through channels of users following forwarded messages.

\begin{figure}
    \centerline{\includegraphics[width=0.49\textwidth]{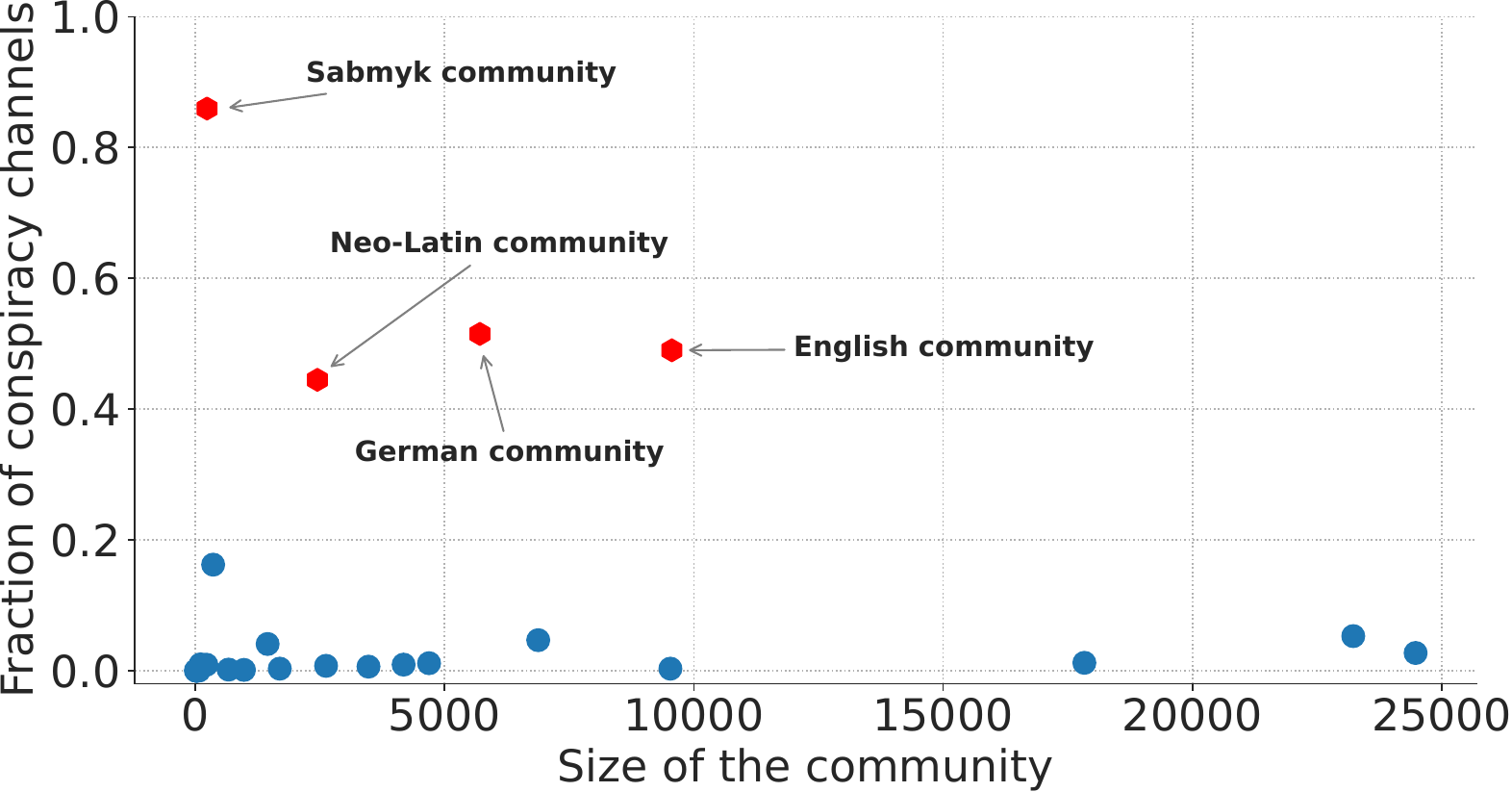}}
    \caption{Each dot represents a community. The y-axis shows the percentage of conspiracy-related channels, and the x-axis represents the community size (number of channels). 
    We highlight in red the communities that show an outstanding amount of conspiracy-related channels.}
  \label{fig:scatterplot}
\end{figure}

Once we have built the graph, we identify communities---subsets of nodes within the graph that are highly connected with respect to the rest of the graph.
In this specific case, a community is a subset of Telegram channels that consistently forward messages among themselves and rarely forward messages from channels of the other communities.
To perform community detection, we use the Leiden algorithm~\cite{traag2019louvain} since it is widely adopted and has proven to be effective in identifying communities in social graphs (\eg Twitter~\cite{sliwa2024case,blekanov2021detection}).
We compute the optimal number of communities with regard to modularity---a metric that measures how much the partitioning is better than a random partition. The metric ranges between  1 to -1, and we obtain a score of 0.78.
With this approach, we identify 47 distinct communities of channels.
Then, we analyze how the conspiracy-related channels are distributed inside these communities.
To illustrate this analysis, we report the scatterplot in Fig.~\ref{fig:scatterplot}. The Figure shows a dot for each community, the number of channels in the community (x-axis), and the percentage of conspiracy channels (y-axis). It is evident that some communities (highlighted in red) stand out due to an unusually high concentration of conspiracy-related channels.
In particular, one community has 86\% of potential conspiracy channels, while the other 3 have more than 40\%. 
These four communities contain 17,950 channels, almost 15\% of all the channels in the TGDataset. 
Furthermore, a vast fraction (78.98\%) of the channels containing at least one link obtained from the Conspiracy Resource Dataset belong to one of these four communities.
In the following, we will refer to the channels in the four communities as Conspiracy Channel Dataset and to all the links in these channels as \textbf{Extended Conspiracy URLs Dataset}. Given the relevance of these communities for understanding the diffusion of conspiracy theories on Telegram, we will analyze them in detail.

\subsubsection{Clustering validation}
\label{sec:clustering_validation}
\reviewed{
As mentioned in the previous section, a substantial portion of channels identified by the clustering algorithm as part of conspiracy communities do not contain links from the Conspiracy Resource Dataset. Since we have less evidence that these channels are conspiracy-related, we manually validate a sample of these channels to evaluate and mitigate the presence of false positives.\\
To perform this task, we select the top 5\% of channels by the number of subscribers, amounting to 414 channels, and perform a manual analysis to assess whether they contain conspiracy theory-related content.
Three researchers independently reviewed and analyzed the content of each channel, annotating the presence of conspiracy-related content. We adopted a conservative approach, flagging only channels explicitly mentioning conspiracy theories, excluding those implying or hinting at them. Furthermore, to minimize the likelihood of false positives, a unanimous agreement among all researchers was required to classify a channel as conspiracy-related.
The analysis shows that 293 channels (70.77\%) are related to conspiracy theories. In the following, we discard the 121 channels not linked to conspiracy theories while we provide a detailed examination of false-positive channels in Sec.~\ref{sec:discussion}}.

\begin{table*}
   \centering
   \small
   \caption{%
       Top channels by authority ranking in each conspiracy community.
  }\label{tab:conspiracy_authories}
   \begin{tabular}{c l l l l}
      \toprule
     \textbf{HITS} &  \textbf{English community} & \textbf{German community} & \textbf{Neo-Latin community} & \textbf{Sabmyk community} \\
      \midrule
    1 & Disclose.tv & Fakten Frieden \#FreeJanich & LA QUINTA COLUMNA TV & sabmyk \\
    2 & Tommy Robinson News & Eva Herman Offiziell & Noticias Rafapal & ChicagoReporter \\
    3 & RT News & Uncut-News.ch "Das Original" & El Investigador.org & GreatAwakeningChannel \\
    4 & Police frequency & Freie Medien & COVID-1984 & CapitolNews \\
    5 & Covid Red Pills  & \#freejanich Oliver Janich öffentlich & DESPERTADOR DE LA MATRIX  & NicolaTeslaNews \\
      \bottomrule
    \end{tabular}
\end{table*}

\subsection{A look into  Conspiracy Communities}
\label{sec:bird-eye-view}
This section offers an overview of the discovered communities, focusing on their most influential channels. We use the Hub and Authorities algorithm (HITS)\cite{borodin2001finding}, originally developed to identify relevant web pages\cite{kleinberg1999hubs}. The algorithm identifies good hubs as those linking to high-authority pages, and good authorities as those linked by many good hubs.

We adapt this idea to the Telegram graph, defining a channel as authoritative if many good hubs forward its messages, and as a good hub if it forwards messages from many authoritative channels. In our analysis, we are particularly interested in highly authoritative channels. Indeed, according to the definition, these channels are very influential in the community as their messages are widely forwarded. Using the HITS algorithm, we identify the top five channels with the highest authority scores in each community, reported in Tab.~\ref{tab:conspiracy_authories}.
Moreover, to better analyze each community, use LangDetect~\cite{langdetect} to identify the languages used in channels. 
In the following, we will analyze separately each community.

\textbf{English community.} This is the largest community, with 9,491 channels and more than 26 million messages sent. More than 89\% of these channels communicate using the English language. The most influential channel of this community is \textit{Disclose.tv}, a website that discusses alternative viewpoints on the news and is notorious for propagating conspiracy theories~\cite{disclosetv}. Following closely in terms of authority ranking is \textit{Tommy Robinson News}, a channel allegedly managed by Tommy Robinson, a British activist known to promote conspiracy theories, particularly those related to the threat of Islam for Western societies~\cite{cleland2020charismatic}. 
\textit{RT News} is a state-funded international media company headquartered in Russia,  known for its alleged bias and for disseminating information that supports the Russian government's positions~\cite{wagnsson2023paperboys}. 
About the \textit{Police frequency} channel, by searching online, we did not find evidence that this channel is associated with well-known entities or individuals linked to conspiracies. However, looking at its messages, we find that it is a far-right channel that focuses on American news about law enforcement, anti-gun control, and anti-immigration.
Finally, \textit{Covid Red Pills} claims to unveil the truth behind the COVID-19 pandemic.

\textbf{German community.} The German community comprises 5,688  channels that share more than 17 million messages. Over 94\% of these channels communicate in German. 
Among the most influential channels, three of them, \textit{Fakten Frieden \#FreeJanich }, \textit{Uncut-News.ch "Das Original"}, and \textit{Freie Medien}, propose themselves as alternative media that share unmanipulated and free news, emphasizing their independence from government or political parties. Instead, the other two channels feature well-known German personalities.
\textit{Eva Herman Offiziell} claims to be the official channel of Eva Herman, a former German news presenter known for promoting various conspiracy theories~\cite{barker2020germany}.
The final channel focuses on Oliver Janich, a prominent supporter of QAnon in Germany~\cite{oliverjanich}, well known for writing conspiracy books about 9/11.

\textbf{Neo-Latin community.} In this case, the community is not predominantly monolingual, with 58\% of the channels primarily communicating in Spanish, 21\% in Portuguese, and 16\% in Italian. This community consists of 2,415 channels and has shared over 8.8 million messages.
Similarly to the German community, the most influential channels present themselves as independent media, advocating freedom of speech and claiming freedom from government influence. Notably, three channels (\textit{El Investigador.org }, \textit{COVID-1984} and \textit{DESPERTADOR DE LA MATRIX}) mostly focus on COVID-19 conspiracies, claiming that the virus is created in a laboratory, the vaccine was created to reduce the population, and the World Health Organization (WHO) is a genocidal organization.

\textbf{Sabmyk community.} The last community stands out, with over 86\% of its 235 channels sharing URLs related to conspiracy theories.
Almost the entire community (95\% of channels) communicates in English.
The authority scores of this community reveal a unique pattern: the \textit{sabmyk} channel is the only true authority, with other channels primarily forwarding its messages. Searching on the web, it emerges that Sabmyk is a complex conspiracy theory proposed as a successor to QAnon~\cite{la2018pretending}. This theory celebrates a new messianic figure called Sabmyk, who actively promotes conspiracy theories against COVID-19 vaccines and concerning the 2020 US elections~\cite{sabmykguardian}.

\subsubsection{Longitudinal analysis}
\begin{figure*}
    \centerline{\includegraphics[width=0.99\textwidth]{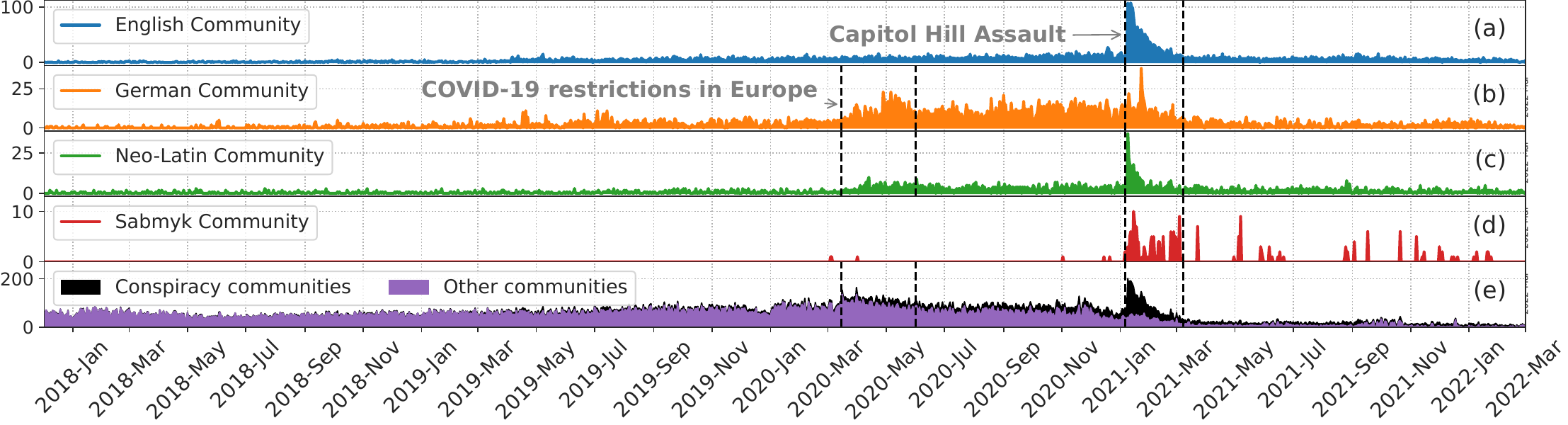}}
    \caption{Channels created on Telegram over time.}
    \label{fig:longitudinal}
\end{figure*}
\label{sec:longitudinal}
An interesting aspect to explore is how conspiracy communities have evolved over time. We analyze this dimension in Fig.~\ref{fig:longitudinal}, which shows the number of channels created daily on Telegram.
The Figure is divided into five parts: the first four charts ($a$,$b$,$c$,$d$) analyze each community, while the last one ($e$) compares the four conspiracy communities aggregated against the other Telegram communities.
The chart shows that the creation of conspiracy channels is not evenly distributed over time. Instead, we find two spikes in channel creation. The first one begins around mid-March 2020, reaches a peak in May, and starts declining until June 2020. Instead, the second spike is much steeper and goes from January 6, 2021 to the middle of March 2021.  

Fig.~\ref{fig:longitudinal} (e) shows a first insight into this phenomenon. The increase in channel creation is more evident in the conspiracy communities (black line) than in the rest of Telegram (purple line). This is particularly evident in the second spike.
This behavior suggests that the spike in conspiracy channel creation is not driven by overall Telegram platform growth but by specific events. Therefore, we examine the messages and descriptions of conspiracy channels created during these periods to understand their origins.
The first peak can be directly linked to the stringent COVID-19 restrictions imposed in Europe during that period and is more prevalent in the German and Neo-Latin communities. We find the surge in Telegram channels offering alternative viewpoints on the pandemic.
Instead, we find that the second peak is linked with the unprecedented Capitol Hill events. During this period, we observed the emergence of several pro-Trump channels, especially in the English community, that became focal points to promote alternative discussions surrounding these events.

\section{Monetization}

The manual review of hundreds of messages from the Conspiracy Channel Dataset during the previous phase revealed that, besides promoting conspiracy theories, some channels post messages to sell products or promote crowdfunding campaigns.
This discovery raises the question of whether some conspiracy channels try to exploit their followers for financial gain.
\reviewed{Intrigued by this aspect, 
we utilized a semi-automatic approach on the URLs shared by conspiracy channels to identify the possible strategies they could use to monetize. First, we extracted domains from the links and determined their frequencies. Then, we used SimilarWeb, a popular web analytics service that categorizes websites and ranks them by traffic, to exclude domains associated with social media and news platforms. Finally, we ordered the remaining domains by frequency. We inspected the remaining domains and searched online to determine whether they could be used to generate revenue for the conspiracy theory channel admin. During these operations, we observed that the inspected websites belong to three main groups: donation, crowdfunding, and affiliate programs.}

\subsection{Donation platforms}
\label{sec:donation_platforms}

\begin{table}
   \centering
   \small
   \caption{%
    Summary of metrics about donation platforms. Gain with (*) indicates monthly earnings.
  }\label{tab:donation}
   \begin{tabular}{l l l l l l}
      \toprule
    \textbf{Platform} & \textbf{URLs} & \textbf{Profiles} & \textbf{Gain (\$)} & \textbf{Donors}  \\
      \midrule
    BuyMeACoffee & 3,157 & 159 & 821,816 & 19,841\\
    Patreon & 31,041 & 936 & 259,585~* & 105,692 \\
    Ko-Fi & 1,475 & 97 & 88,209 & 5,502  \\
    SubScribeStar & 2,604 & 134 & 5,947~* & 5,675\\
    Paypal/donate & 83,512 & 1,434 & -  & - \\
    DonorBox & 1,550 & 59 & - & - \\
    \midrule
    \textbf{Total} & 123,339 & 2,819 & 1,175,557 & 136,710 \\
      \bottomrule
    \end{tabular}
\end{table}

The first monetization strategy we analyze consists of asking the subscribers for donations to support the channel and its activity. 
As a first step, we identify the most popular donation platforms that can be used for this purpose.
We manually collect them by analyzing the results of Google queries such as: \textit{top donation platforms}. 
Given the presence of language-specific communities, we conducted country-specific queries containing different languages and keywords. 
Moreover, to avoid search results on Google from being affected by our browsing history or geo-location, we conduct queries using a VPN. 
In this way, we collect 31 donation services.
Then, we analyze the presence of donation platforms in the URLs shared in the Extended Conspiracy URLs Dataset.

During the exploration phase, we discover a few cases of messages sharing donation URLs with the intent to discredit genuine content creators' donation campaigns rather than promote them. 
\reviewed{To discard these cases, we analyze the content of the messages to assess the intent of the authors.
For this task, we utilize the GPT-4o model via the ChatGPT API, asking the LLM to infer whether the message's author is requesting funds or is discrediting a donation project.
In the Appendix, Tab.~\ref{tab:chatgpt_prompt} reports the prompt we provided to the LLM.
Of the 56,804 unique messages containing donation links, the model identifies 56,532 messages (99.52\%) as promoting a donation page, while 272 are against the analyzed campaign.
To evaluate the performance of the the model, we randomly sample 200 messages, 100 marked as positive (\ie the author is promoting a donation page) and 100 as negative (\ie the author is discrediting the linked donation campaign). Then, we manually label each message, obtaining a ground truth of 186 positive messages and 14 negative. Upon evaluating the model's performance, we observed that it excels at detecting negative messages, correctly classifying all 14 negative instances.}
\reviewed{However, 86 messages that initially the model classified as negative are actually positive. In other words, the model may classify promoting campaigns as messages discrediting them. Further investigation revealed that this issue is particularly evident in channels that consistently append donation links at the end of all their messages. In these cases, the model may incorrectly assume that messages with a negative tone aim to discredit the link, even when there is no actual relationship between the message content and the link. Nevertheless, considering our goal to exclude discrediting messages and the relatively small number of potentially misclassified messages (272 - 0.48\%), we find the model’s performance satisfactory for our purpose.}

After filtering out messages classified by the LLM as discrediting the linked donation page, we identified 124,610 URLs from 15 services, shared across 5,585 channels, accounting for 31.3\% of our dataset.
The most used platforms by number of URLs are Paypal/donate~\cite{paypaldonate} with 83,512 URLs (67\% of the donation URLs), Patreon with 31,041 URLs (24.9\% of the donation URLs), BuyMeACoffee with 3,157 (2.5\%), SubScribeStar with 2,604 (2.1\%), DonorBox with 1,550 (1.2\%) and Ko-Fi with 1,475 (1.2\%).

To estimate the potential earnings of the Conspiracy Channel Dataset on these platforms, we create a custom crawler using the Selenium Framework~\cite{selenium}. We crawl all the URLs from the 6 most used donation platforms in our dataset, covering 98.9\% of the URLs\footnote{\reviewed{We conducted the crawling of donation, crowdfunding, and affiliate URLs in September 2023.}}.
Unfortunately, PayPal/donate and DonorBox do not provide information about donation amounts and the number of donors. Thus, by scraping the web pages of these services, we could only determine the number of unique profiles still reachable.
Instead, we obtained information regarding the number of donors and donated amounts for most profiles on the other services.
Specifically, for Patreon and SubscribeStar, we retrieve the number of donors and the total amount of money donated in the last month. As for BuyMeACoffee and Ko-Fi, we collect data on the number of donors and the total amount of money raised by the campaigns' creators.

Tab~\ref{tab:donation} shows the number of URLs we find for the six most used services, the number of created campaigns, the amount of money donated, and the number of donors. 
Among the profiles we examined, the one that raised the most money through this strategy is \textit{QAnon Anonymous}\cite{QAnonanonymous}, which received an astonishing \$87,955 from 20,103 subscribers in September 2023 using Patreon.

\begin{table}
   \centering
   \small
   \caption{%
    Summary of metrics about crowdfunding platforms.
  }\label{tab:crowfunding_summary}
   \begin{tabular}{l r r r r}
      \toprule
    \textbf{Platform} & \textbf{URLs} & \textbf{Projects} & \textbf{Funds (\$)} & \textbf{Backers}  \\
      \midrule
    Givesendgo & 14,431 & 607 & 35,228,534 & 519,515\\
    GoFundMe & 4,314 & 1,070 & 27,688,574 & 284,802\\
    Kickstarter & 234 & 64 & 3,491,996 & 39,112 \\
    Fundrazr & 184 & 36 & 2,270,651 & 32,981 \\
    Indiegogo & 53 & 32 & 1,061,501 & 7,207 \\
    Fundly & 299 & 18 & 717,714  & 14,229 \\
    DonorBox & 1,073 & 30 & 191,463 & 4,276 \\
    Paypal/pools & 4,082 & 392 & 115,455  & 3,513 \\
    \midrule
    \textbf{Total} & 24,670 & 2,249 & 70,765,888 & 905,635 \\
      \bottomrule
    \end{tabular}
\end{table}


\subsection{Crowdfunding and Fundraising services}
\label{sec:crowd}
Conspiracy channels can ask their followers to support them on a specific project by leveraging crowdfunding services. In these platforms, individuals and organizations can raise funds for projects or causes by receiving contributions, typically in exchange for rewards or
incentives.
We manually build a list of popular crowdfunding services performing Google queries such as \textit{best crowdfunding services}, following the same methodology used for donation platforms. At the end of the process, we collect a list of 49 crowdfunding or fundraising websites. 
Looking for URLs containing the domain of these platforms in the Extended Conspiracy URLs Dataset, as for the donation links, we find evidence that conspiracy channels occasionally run misinformation campaigns against genuine projects.

\reviewed{Thus, we use the same approach as the previous subsection to discover these cases (Sec.~\ref{sec:donation_platforms}), asking the GPT-4o model to determine the author's intention behind each message (prompt in Appendix, Tab.~\ref{tab:chatgpt_prompt}). Out of 7,499 distinct messages containing a URL to a crowdfunding project, the model classifies 7,319 messages (97.6\%) as promoting campaigns while 180 as messages against the linked campaign. Also, in this case, to validate the model, we randomly sampled 200 messages, 100 marked as positive and 100 as negative.
Then, we manually label these messages as previously done in Sec.~\ref{sec:donation_platforms}. Here, we identified 145 positive samples and 55 negative samples. 
Comparing the model results with the manually annotated labels, we observed the model tends to classify positive messages as negative, similar to the previous task. Specifically, the model incorrectly classified 45 messages endorsing crowdfunding campaigns as negative.
After reviewing the misclassified messages, we find many authors initially discredit the GoFundMe platform, encouraging their audience not to use it, and then endorse a crowdfunding campaign on the GiveSendGo platform. This pattern was common among American far-right channels, which frequently complain about GoFundMe blocking their projects.
Given that the model provides a conservative estimation and considering the relatively low number of misclassified messages (180), we used this model for our classification. Accordingly, we discarded 180 unique messages (357 including duplicates) linking to crowdfunding platforms.}


\reviewed{
As further validation, we manually examine the crowdfunding campaigns that raised more than \$200K, counting 98 different projects, to determine if they are related to conspiracy theories. Three researchers independently analyzed these projects, and a campaign was labeled as related to conspiracy theories only if there was unanimous agreement. 
Among the 107 campaigns, we identified 51 projects (47.66\%) that, despite being supported by conspiracy channels, have goals unrelated to conspiracy theories.
}

After this validation process, we find 24,696 URLs from 18 different platforms, shared by 3,469 channels, covering 19.5\% of our dataset.
The most used services are GiveSendGo with 14,431 URLs (58.4\% of the crowdfunding URLs), followed by GoFundMe with 4,314 URLs (17.5\%), Paypal/pools with 4,082 (16.5\%), and DonorBox with 1,073 (4.3\%).

Also in this case, we implemented a scraper using the Selenium Framework for each of the aforementioned platforms, covering 99.7\% of all the crowdfunding URLs.
This scraper allows us to collect information about each campaign's earnings and analyze their status.
Indeed, a campaign can be completed or ongoing. Kickstarter enables creators to access the funds only if they reach a predefined target funding at the end of the campaign. Instead, on other platforms, the campaign creator can access the funds raised while the campaign is ongoing. An exception is Indiegogo, which allows creators to choose between the two options when starting a campaign.
Upon examining the campaigns' statuses, we discover eight campaigns on Kickstarter and one on Indiegogo that concluded without reaching their fundraising goal. Thus, we do not include the funds raised from these campaigns.

Tab.~\ref{tab:crowfunding_summary} reports the number of URLs, different projects, money raised, and number of backers for the top eight services by number of URLs. As it is possible to note, this strategy is the most remunerative, collectively funding conspiracy theorists with almost \$71M. 
Analyzing the content of the campaigns and find that they typically fall into these categories:

\textbf{Campaigns supporting far-right.} 
We identified several fundraising campaigns for far-right projects, mainly raising funds for legal costs related to the January 6th Capitol riot. Three of these campaigns alone ~\cite{capitolhill1,capitolhill2,capitolhill3} collected \$188,215 from 2,920 backers. We find that many of these campaigns are hosted on Givesendgo, a platform well known for promoting extremist content~\cite{visser2024crowdfunding}.


\textbf{Campaigns about COVID-19.} 
We also find GiveSendGo campaigns frequently linked to COVID-19, including those for the Freedom Convoy 2022~\cite{gordon2022freedom}, a protest movement against COVID-19 policies. Similar GoFundMe campaigns redirect to refund pages due to the platform's ban on content promoting violence~\cite{freedomgofundme}. Looking at the news, we discover that one of these campaigns was the biggest gainer of the platform, raising \$9,748,168 from 112,943 donors~\cite{gofundmeconvoy}. Additionally, Indiegogo and Kickstarter host projects funding documentaries about COVID-19~\cite{coronadoc1,coronadoc2,coronadoc3}. These platforms have stricter content policies, but moderating pseudo-scientific claims remains challenging due to freedom of expression concerns.


\textbf{Scam campaigns.} We detect crowdfunding campaigns that are outright scams. An example is a Fundrazor campaign~\cite{scamcampaign1} falsely claiming to raise funds for starving children in Venezuela. It collected \$39,500 from 602 donors before being shut down, and a banner on the page states that Save The Children has confirmed they are not associated. Another scam~\cite{scamcampaign2} on Indiegogo raised over \$7.4K for a portable air cleaner claiming to purify the air with a high-frequency generator before being closed as the creator claimed that Silicon Valley internet companies had hindered the project.

\textbf{Campaigns against the establishment.}
We found campaigns raising funds to challenge government policies or influential figures. For example, a campaign~\cite{helpkids} funded by American Education Defenders, Inc., aims to counter U.S. school indoctrination with a program called "America's 52 Videos" that promotes American values. Another campaign~\cite{pyramidpower} supports a documentary series on global manipulation. Together, these campaigns raised over \$37K from 477 contributors.

\subsection{Affiliate Programs}
\label{sec:affiliation}

\begin{table}
   \centering
   \small
   \caption{%
    Summary of metrics about e-commerce platforms.
  }\label{tab:ecommerce}
   \begin{tabular}{l l l l l l}
      \toprule
    \textbf{Platform} & \textbf{URLs} & \textbf{Products} & \textbf{Affiliate} \\
    \midrule
    Amazon      & 63,445 & 19,372 & 34,412 \\
    Teespring   & 2,415 & 336 & - \\
    eBay        & 1,932 & 958 & 131 \\
    Etsy        & 1,082 & 512 & - \\
    \midrule
    \textbf{Total} & 68,874 & 21,178 & 34,543  \\
      \bottomrule
    \end{tabular}
\end{table}

Conspiracy channels adopting this strategy use affiliate programs from e-commerce platforms to earn a commission.
Thus, as a first step, we identify the most popular e-commerce platforms they can utilize for this purpose. To this end, we leverage SimilarWeb to retrieve all the 42 services reported in the global rank for the \textit{eCommerce \& Shopping} category.
Then, we extract URLs containing the domain of these platforms from the Extended Conspiracy URLs Dataset, finding 70,149 URLs from 28 platforms.
Among the top 4 platforms, covering 98.1\% of e-commerce links, only Amazon and eBay offer an affiliate program.
In this program, a participant, known as partner, can generate unique links pointing to products on the platform that embed his unique identification number. The partner earns a commission for any purchases made by users who land on the platform through his link.
In particular, Amazon's partners earn a commission, between 1\% and 12\% depending on the categories of the product and the location of the targeted market (\eg amazon.com, amazon.de) from all the items bought in the next 24 hours from arrival through the link~\cite{amazonaffiliate}. Similarly, on eBay, a partner earns a commission between 1\% and 4\% on the items purchased on the platform in the following 24 hours~\cite{ebayaffiliate}.

Overall, we find 63,445 URLs pointing to Amazon and 1,932 to eBay in the Conspiracy Channel Dataset.
To detect URLs belonging to the affiliate program, we search for URLs containing the parameters \textit{tag=} for Amazon and \textit{campid=} for eBay.
Surprisingly, we discover that 34,412 (54.2\%) of Amazon's URLs and 131 (6.8\%) on eBay are affiliated links granting commissions to the channels and that 1,810 channels (10.2\% of our dataset) leverage this strategy. Tab.~\ref{tab:ecommerce} provides a summary of the metrics of e-commerce platforms.

Unfortunately, since there is no public information available about the partner programs, we can not estimate the gain of the channels with this strategy. 
However, we use the Selenium Framework to build a custom crawler to retrieve the type of goods sold on this platform.
We obtain the item category for 50,295 of the 63,445 Amazon URLs, representing 17,163 
different products. 
\reviewed{Interestingly, we discover that more than half of the advertised products are books. The remaining goods fall into various categories, such as Electronic Devices (\eg, torches, GPS devices, power supplies) and Health \& Personal Care (\eg, vitamins, proteins, disinfectants). These items are marketed as essential for surviving in post-apocalyptic scenarios caused by the side effects of vaccines and 5G radiation. Tab.~\ref{tab:amazon_products} in the appendix reports the number of URLs and the items found for each category. Instead, in Sec.~\ref{sec:discussion}, we discuss the danger behind promoting these goods and the reason for the significant presence of books among the links.}


\subsection{Monetization Analysis}
\label{sec:monetization_analysis}

\begin{table}
   \centering
   \small
   \caption{
    Number of channels for each monetization strategy.
  }\label{tab:monetization_community}
   \begin{tabular}{l l l l}
      \toprule
      \textbf{Community} & \textbf{Donation} & \textbf{Crowdfunding} & \textbf{Affiliate}  \\
      \midrule
    English &  2,086 & 1,930 & 390  \\
    German & 2,769 & 1,233 & 1,224 \\
    Neo-Latin & 728 & 305 & 168  \\
    Sabmyk & 2 & 1 & 28 \\
    \midrule
    \textbf{Total} & 5,585 &  3,469 & 1,810 \\
      \bottomrule
    \end{tabular}
\end{table}


\reviewed{Tab.~\ref{tab:monetization_community} shows the number of channels in each community that use the three monetization strategies.
As we can see, most German and Neo-Latin channels predominantly use the donation strategy rather than crowdfunding and affiliate programs. In contrast, English channels use crowdfunding significantly more frequently than other communities.
This observation is consistent with previous analyses showing that crowdfunding campaigns often concentrate on events like the Capitol Hill Riot and the Freedom Convoy, arguably more significant for English-speaking communities.
Moreover, German channels use the affiliate strategy more commonly than other communities. This phenomenon is largely explained by a specific group of channels we will explore in the following paragraph.
Finally, we find that Sabmyk shares almost no monetization links. A closer examination reveals that Sabmyk primarily monetizes by directing users to the website of the creator of the movement to promote and sell his artworks. Analyzing the affiliate links promoted within the Sabmyk community, we find they are shared across 28 channels using the same referral ID, highlighting a high level of coordination.}

\begin{figure}
    \centerline{\includegraphics[width=0.45\textwidth]{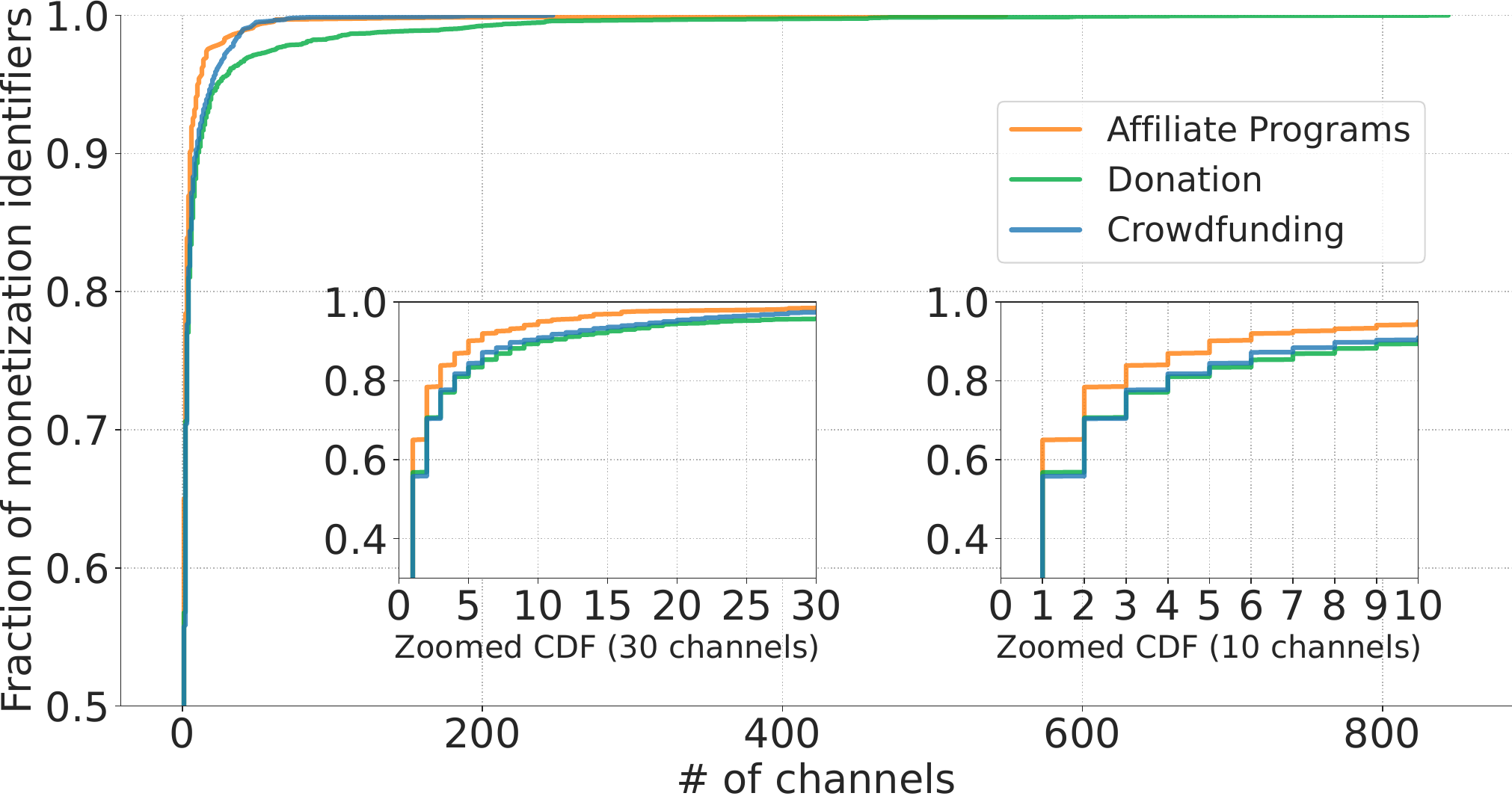}}
    \caption{Distribution monetization projects in channels.}
  \label{fig:cdf_multiple_channels}
\end{figure}

\reviewed{\textbf{Monetization links promoted by multiple channels.}
The analysis of Sabmyk affiliate links highlights the sharing of the same referral link by multiple channels. To study this phenomenon in more depth, we conducted a broader analysis, including donations and crowdfunding projects.
Fig.~\ref{fig:cdf_multiple_channels} shows the cumulative distribution function (CDF) depicting the fraction of referral IDs (orange), crowdfunding (blue), and donation projects (green) shared in channels. The main plot provides an overview of the entire distribution, while two subplots show two different levels of zoom.
By examining the zoomed CDFs, we can observe that most referral links, donation projects, and crowdfunding projects are typically shared by only a few channels. Specifically, we find that 65.4\% of affiliate IDs, 56.8\% of donation projects, and 55.8\% of crowdfunding projects are shared by only a single channel. The figure also highlights that affiliate links are shared by fewer channels than donation and crowdfunding projects, with almost 80\% of referral links shared by at most two channels.
This can be expected as, while affiliate links are primarily beneficial to the channel owner, crowdfunding and donation projects may sometimes support broader causes and may be shared by channel owners even if they are not the direct beneficiaries of the funds received.
Looking at the tail of the overall distribution, we find other interesting insights.
Examining crowdfunding projects, we find that the most widely shared is a highly popular project supporting the Freedom Convoy, promoted on GiveSendGo~\cite{freedomconvoy}. This project is shared by 247 conspiracy channels and has raised \$9.7 million.
For donation projects, we identified a donation link that was shared by 844 channels. This link is associated with the \textit{Corona\_Fakten} channel, which focuses on disseminating news and controversial theories about COVID-19. The channel frequently includes the donation link in its posts, and because its content is widely shared and forwarded by many other channels, the donation link has spread across hundreds of channels.
What is more surprising instead, is the presence of a group of over 549 channels promoting Amazon products using the same affiliate ID. Our analysis reveals this group of channels is entirely located in the German community and is actively involved in promoting various conspiracy theories, predominantly related to COVID-19.}


\begin{table}
   \centering
   \small
   \caption{%
    Monetization strategies used by channels.
  }\label{tab:monetization_strategies_multiple_channels}
   \begin{tabular}{l l l l }
    \toprule
    {} & \multicolumn{3}{c}{\textbf{\# of monetization strategies used by channels}} 
    \\ \cmidrule{2-4}

    \textbf{Community} & \multicolumn{1}{c}{0} & \multicolumn{1}{c}{1} & \multicolumn{1}{c}{2+}  \\
      \midrule
    English & 80.9\% (7,680) & 15\% (1,426) & 4.1\% (385) \\
    German & 73.8\% (4,196) & 20.9\% (1,187) & 5.4\% (305) \\
    Neo-Latin & 79.0\% (1,909) & 17.1\% (414) & 3.8\% (92) \\
    Sabmyk & 99.6\% (234) & 0.4\% (1)  & 0\% (0)  \\
    \midrule
    \textbf{Total} & 78.6\% (14,019) & 17\% (3,028) & 4.4\% (782) \\
      \bottomrule
    \end{tabular}
\end{table}

\reviewed{\textbf{Channels using multiple monetization strategies.}
Finally, Tab.~\ref{tab:monetization_strategies_multiple_channels} shows the number of monetization strategies used by channels divided by community.
We exclude from this analysis links contained in forwarded messages to avoid the cases, of news websites reporting donation links that are extremely widespread and are instead not a monetization strategy of the considered channel.
Interestingly, we find that the majority of channels in all the communities do not use monetization strategies.
Among the channels that monetize, most use only one strategy, with 4.4\% employing two or more strategies.
}

\reviewed{\textbf{Monetization trends over time}}
\label{sec:monetization_over_time}
\begin{figure}
    \centerline{\includegraphics[width=0.45\textwidth]{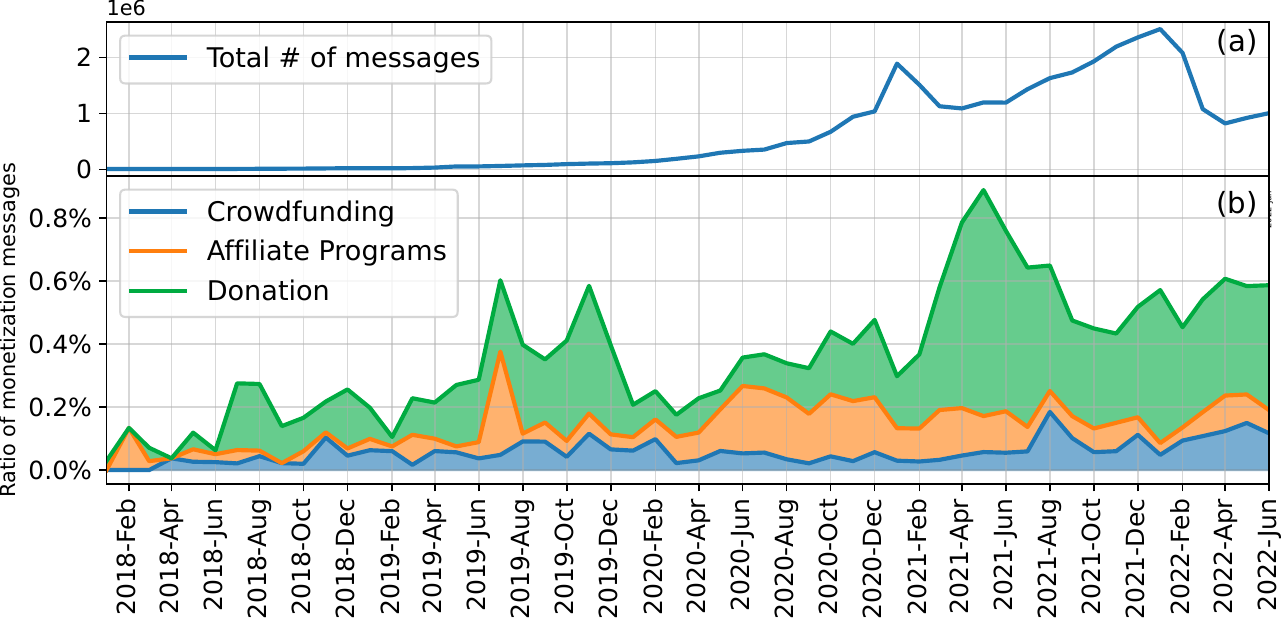}}
    \caption{Total number of messages and fraction of monetization messages over time.}
  \label{fig:monetization_over_time}
\end{figure}
\reviewed{Fig.~\ref{fig:monetization_over_time} shows the total amount of messages sent by conspiracy channels (a) and which percentage of these messages are related to monetization (b). Monetization messages are divided according to the type of monetization in crowdfunding (blue), affiliate programs (orange), and donation (green). We observe that monetization messages are generally not extremely common, reaching a maximum of 0.8\% in early January 2021.
However, the phenomenon has clearly evolved from barely existing in 2018 to achieving relative stability of almost 0.4\% of all messages in 2022. From chart (a), we can notice that the total number steadily decreased after January 2022, but instead, the ratio of monetization messages remains stable. 
It is also interesting to note that monetization strategies were almost equally divided before 2021, but then the donation strategy became much more prevalent. A possible explanation could be that crowdfunding platforms have implemented increasingly stricter policies to counter questionable projects, an aspect that we will explore in the following paragraph.}

\reviewed{\textbf{Shift in platform usage.} Fig.~\ref{fig:platform_usage_over_time_donation} and Fig.~\ref{fig:platform_usage_over_time_crowdfunding}, in the Appendix, show, respectively, the proportion of donation and crowdfunding platforms used by each community monthly. Examining donation platforms, we observe distinct preferences across different linguistic communities. The German community predominantly uses PayPal. In contrast, Patreon has been the most used platform in the Neo-Latin community over the last year. The English community initially used Patreon and PayPal most frequently, but in the past 18 months, there has been a significant diversification, with multiple platforms being utilized.
For crowdfunding, PayPal Pools was the dominant platform among both the German and Neo-Latin communities. However, its popularity decreased until it was discontinued in September 2021 and replaced by GoFundMe. 
The English community displayed a different trend. GoFundMe was consistently the preferred platform, as PayPal Pools was never widely adopted. Around October 2020, GiveSendGo began to rise in popularity, accounting for 75\% to 90\% of all URLs shared by the English community monthly. This shift may be attributed to GoFundMe stricter policies, as GiveSendGo promotes itself as having no moderation. We find evidence of this transition in the messages of the English community.}

\subsection{Other Strategies}
\label{sec:other_strategies}
In addition to the monetization strategies analyzed previously, conspiracy channels can exploit other sources of revenue.

\textbf{Other Amazon features.}
The first one is related to the Amazon Influencer Program~\cite{amazoninfluencer}. This program enables Influencers to create an Amazon web page with some selected products, earning commissions on sales.
These pages are easily detectable because they contain the \textit{/shop/} string in their URL. Looking for this pattern in our dataset, we find 25 different shops.
Another possibility is using Amazon's wish lists, essentially lists of products a user desires. These lists can be private (visible only to the creator) or public. In the latter case, anyone with access to the wish list link can gift a product to the list's creator.
To identify wish lists in our dataset, we extract URLs containing the \textit{/wishlist/} string in their path. This process led us to 119 URLs, pointing to 32 distinct public lists. The inspection of these lists revealed that 12 are no longer accessible, and the others contain a wide variety of products, including underwear, vitamins, survival kits, and prepaid cards.

\textbf{Blockchain addresses.} In our analysis, we discover conspiracy channels asking for cryptocurrency donations in their descriptions. 
We use regular expressions to extract wallet addresses of the most popular blockchains (Bitcoin and Ethereum), as well as the prominent privacy-preserving blockchains (Monero and Zcash). We identified these blockchains' addresses in 40 channels, and through manual verification, we confirmed that 29 of them are used for donations. Analyzing the BTC wallets, we find they received 115 transactions, totaling 0.5 BTC ($\approx\$13,000$). Performing the same analysis on Ethereum, we find that the wallets received 42 transactions, amounting to 5.5 ETH ($\approx\$9,000$).

\textbf{Custom websites.} 
Analyzing the URLs and examining messages within the conspiracy channels, we observed frequent promotion of custom e-commerce sites or personal websites. Inspecting these websites, we discover that many of them feature dedicated donation sections with blockchain addresses or various payment options.
Thus, we perform a raw analysis to estimate the magnitude of the phenomenon. In particular, we look for URLs containing the words: \textit{shop, products, store, produkt, collections, donate, donations or support} as third-level domains or that have these words in the URL's path. As a result, we find 152,680 URLs (24,980 unique) matching our definition.
Unfortunately, we can not validate or analyze this huge amount of websites since it requires a heavy manual effort or to build custom parsers.

\textbf{Drive traffic to video hosting services.} The URLs analysis also revealed that conspiracy channels share a considerable number of links to popular video hosting services such as YouTube (3,656,171 URLs) or BitChute~\cite{trujillo2020bitchute} (276,513 URLs).
While some channels may share videos as a resource to confirm their theories, it is also well known from previous work~\cite{ballard2022conspiracy}, that some of them leverage such platforms to monetize their content through Partner Programs (\eg YouTube Partner Program~\cite{cheng2014understanding}). These programs allow content creators to earn money by placing advertisements in their videos and paying them proportionally to the time they are viewed. While the in-depth analysis of this phenomenon falls outside the scope of this work, we believe that the dataset we release can be a valuable resource for future research in this area.

\textbf{Channel ads.} Finally, conspiracy channel administrators could also monetize by publishing sponsored messages to their subscribers. There are mainly two methods to implement this strategy. 
The first relies on a feature recently introduced on Telegram, the Sponsored Messages~\cite{telegramads}. This functionality enables channel owners to share sponsored messages to receive a share of the advertising revenue.
However, it is worth noting that this feature is relatively new and still in the beta phase. 
The second approach involves using external services like \textit{telega.io}, which act as intermediaries between channel administrators and advertisers or establish private deals directly between advertisers and channel administrators.
However, this kind of sponsored message is likely impossible to detect when products are deceptively promoted into the content and storyline of the channel.

\section{Discussion}
\label{sec:discussion}
\reviewed{\textbf{Impact of the time gap.}
Messages in the TGDataset range from October 2015 to July 2022, while we retrieved monetization information between September and October 2023. We analyzed the impact of this gap to determine if it could affect our estimation of money raised.
Concerning the crowdfunding platforms Kickstarter, Indiegogo, and PayPal/pool, the time gap has no impact. Kickstarter and Indiegogo campaigns can run for up to 60 days, while PayPal/pool, discontinued in November 2021, allowed campaigns to run for only 30 days.
Conversely, projects on other platforms can run indefinitely. 
Therefore, we examine the distribution of donations over time by extracting all available donations from platforms that run campaigns without time constraints.
From GiveSendGo, we retrieved complete donation data, while on GoFundMe, we had access only to the most recent 1,000 donations and the top 1,000 highest donations for each project. Finally, we could not extract Fundrazr data, since the platform hides the amount users contributed. While these limitations prevent us from obtaining a complete picture, the available data still offer valuable insights. 
Our analysis revealed that 76.3\% of campaigns closed or had their last donation by July 2022.
Moreover, we found that, on average, each campaign raises 34.2\% of its total funds within the first 7 days, 65\% after one month, and 74.4\% after three months. Therefore, since most of the money is raised within the first three months from the launch, the time gap has a limited impact.}
\\
\reviewed{\textbf{False positive channels.}
Validating channels present in the detected communities Sec.~\ref{sec:clustering_validation}, we 
classify 121 out of 414 channels as false positives.
In particular, we identified 28 channels associated with the American far-right or supporting the former Brazilian president Bolsonaro. Although the messages on these channels do not explicitly endorse or promote conspiracy theories, they often share a similar underlying narrative.  Additionally, the relationship between conspiracy theories and the American far-right~\cite{us_far_right_qanon} or Bolsonaro~\cite{demuru2020conspiracy} is well established.
Furthermore, 24 channels cover topics related to religion and self-improvement. According to social and psychological scientific works~\cite{douglas2017psychology, bond2023rise}, there exists a connection between these topics and conspiracy theories. Indeed, conspiracy theories often provide a sense of belonging to a community and the perception of control over complex and seemingly chaotic events. Moreover, some conspiracy theories, like QAnon, incorporate seemingly religious elements such as beliefs in a secretive satanic elite, apocalyptic prophecies, and the involvement of divine or demonic forces.
Surprisingly, 15 channels are linked to cryptocurrency. This seemingly unusual connection is backed by a recent study, which reveals that cryptocurrency holders often exhibit strong anti-establishment sentiments and tend to believe in conspiracy theories~\cite{littrell2024political}. Lastly, there are 13 channels dedicated to alternative news. These channels provide viewpoints that often challenge traditional narratives.}
\\
\reviewed{\textbf{Detecting new monetization techniques.}
In this work, we focus on three main monetization methods used by conspiracy channels and mention other methods we've observed. However, new techniques may appear in the future. Based on our experience, we suggest the following ways to identify potential new techniques.
Using the monetization messages collected in this study, an LLM can be fine-tuned to detect these messages more effectively than the standard ChatGPT we used. By monitoring the links these channels share and analyzing the most frequently shared ones with the fine-tuned model, it will be possible to identify new platforms used for monetization.
Finally, many monetization strategies used by conspiracy channels are similar to those used by web content creators. Therefore, if they adopt new strategies, it is worth checking if conspiracy channels start using them.}

\textbf{The risks of conspiracy narratives.}
Our analysis reveals a clear distinction between the products promoted by conspiracy theory channels on Amazon and other online marketplaces.
Indeed, while Amazon features standard products like books, masks, and water filters, we find a range of questionable products (\eg 5G shields, EMF stone protectors, and healing wands) on eBay, Etsy, and Teespring.
The distinction is likely attributed to the different content policies of these platforms.
Indeed, Amazon upholds a more rigorous content policy compared to the other services.
Nevertheless, it is important to emphasize that these products are not inherently harmful. Indeed, the concern lies in the narratives and promotions associated with these items.
For example, ordinary substances like Sodium Chloride and Chlorine Dioxide,  commonly found in pharmacies, are promoted in four channels as essential components to prepare the so-called "Miracle Mineral Supplement", which is reported as a miraculous cure for various diseases, including cancer and HIV.
Similarly, we discover 35 channels sharing links to buy seemingly innocuous white pine needles at an exorbitant price of \$150 on Etsy. The concern lies in the accompanying message, which presents a guide for COVID-19 survival. This guide discourages seeking medical care in hospitals and suggests homemade remedies, including tea prepared with the costly pine needles mentioned above.

\begin{figure}
    \centerline{\includegraphics[width=0.4\textwidth]{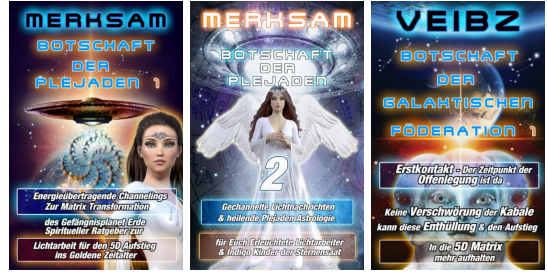}}
    \caption{The covers of the three most shared books by conspiracy channels available for sale on Amazon.}
  \label{fig:conspiracy_books}
\end{figure}

\textbf{Marketplace policies.}
Although Amazon has a rigorous policy about the items sold and actively operates to ban QAnon merchandising~\cite{ballard2022conspiracy}, it has a less strict policy about book content.
Quoting Amazon policies: \textit{As a bookseller, we believe that providing access to the written word is important, including content that may be considered objectionable.}~\cite{amazonbooks}.
The combination of this less strict policy and the simplicity of self-publishing on Amazon allowed conspiracy theorists to spread their ideas and monetize through book sales.
Indeed, nearly 50\% of the links point to books, and almost half of them include an affiliate tag.
Looking at the most frequently occurring authors of books promoted using non-affiliate links, we discover three relatively obscure German writers: VEIBZ (3,091 occurrences), MERKSAM (1,492 occurrences), and EBURD (813 occurrences).
Their books explore several conspiracies with the goal of revealing the truth on subjects such as: "NASA \& Elon Musk – They lie \& cover-up," "CERN \& its satanic roots," "HAARP \& CERN use Alien Tech." Fig.~\ref{fig:conspiracy_books} shows the cover of the three most shared books.


Concerning the promoted items, there are also problems related to the transparency of the activity and compliance with the referral programs.
Indeed, according to the partnership agreement of Amazon and Ebay~\cite{amazonaffiliate}~\cite{ebayaffiliate}, a
partner---a participant that can generate referral links and earn commissions---has to clearly disclose their partnership.
Moreover, near each affiliated link should be a disclosure such as \textit{"(paid link)," "\#ad," or "\#CommissionsEarned"}.
This informs customers that there is a conflict of interest in the promoted item.

\section{Mitigation}
\reviewed{To address the concerns highlighted in the previous section, we developed two tools: The Channel Checker Bot and the ConspiracyAlert plugin. These tools assist users in identifying channels that spread questionable content, information potentially harmful to their health, and crowdfunding campaigns that may support conspiracy theory groups.}

\begin{figure}
\centerline{\includegraphics[width=0.3\textwidth]{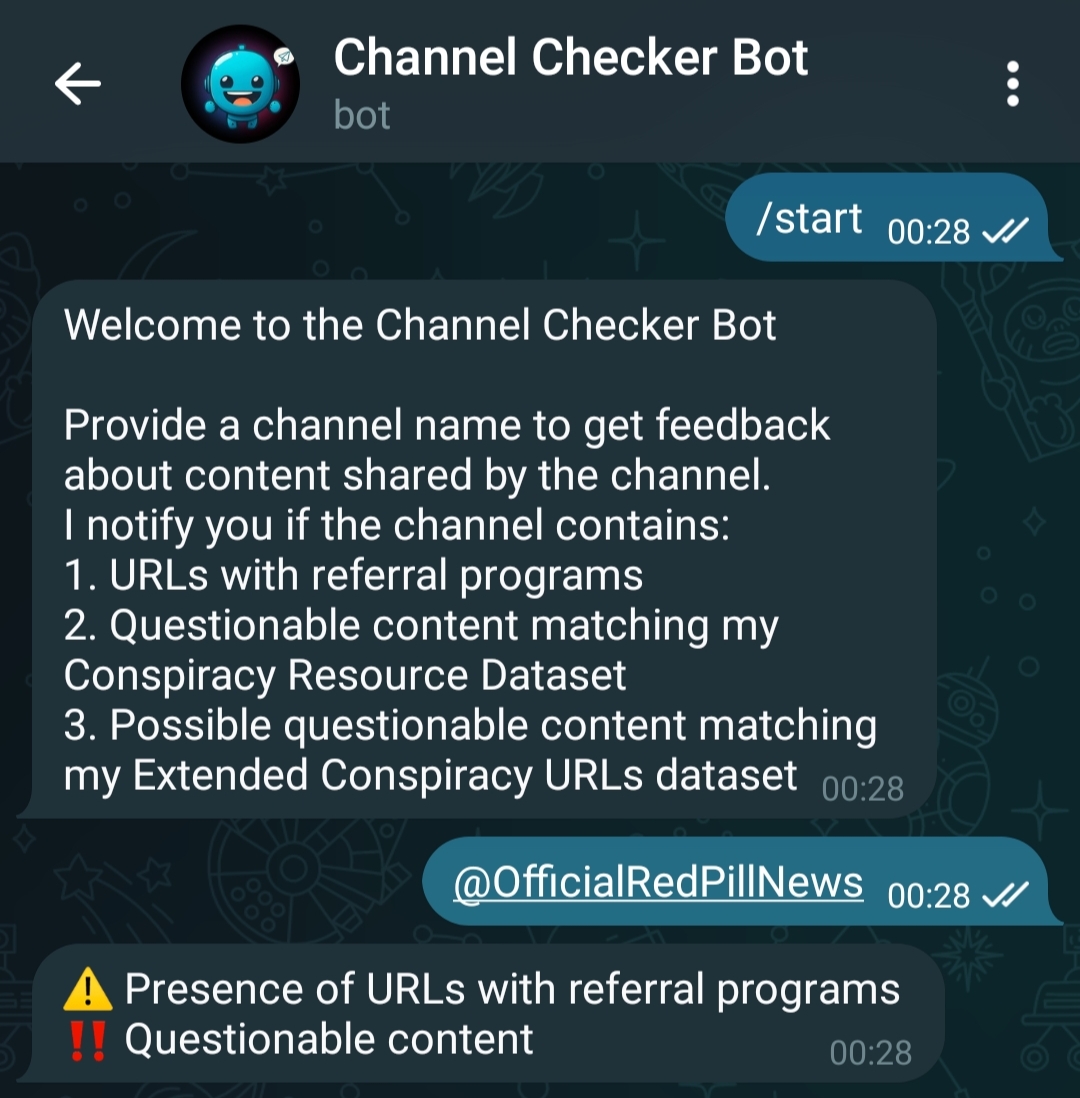}}
    \caption{An instance of the Channel Checker Bot chat.}
  \label{fig:bot_chat}
\end{figure}

\reviewed{\textbf{The Channel Checker Bot.} It is a Telegram Bot implemented in Python. It receives a channel name from the user and provides detailed information about that channel.
The bot uses a database containing channels identified during our research and information such as questionable content, advertised products, and dubious fundraising campaigns. When a user inputs a channel name, the bot searches the database for a match. If the channel is found, the bot provides the user with the existing information from the database.
Instead, if the channel is not in the database, the bot initiates a headless Telegram client, joins the channel via Telegram APIs, and retrieves all shared content. Then, it extracts all shared links and checks them against the conspiracy resource dataset and the monetization links collected during this research. After the analysis, the bot updates the database with the new channel and its associated information. Finally, the bot presents the analysis result to the user. Fig.~\ref{fig:bot_chat} displays an instance of a chat with the Channel Checker Bot.}
\reviewed{Therefore, the Channel Checker Bot serves a dual purpose: It provides users with detailed information about channels and continuously updates the database with new conspiracy channels.}

\reviewed{\textbf{ConspiracyAlert: A Browser Plug-in.}
Although the Channel Checker Bot serves as a warning system for Telegram users navigating the platform, it does not help users who browse the web and land on questionable products or crowdfunding campaigns. To address this gap, we developed the ConspiracyAlert browser plug-in.
This plug-in, compatible with Chrome and written in TypeScript, monitors the user's web navigation. Once a webpage is fully loaded, the plug-in captures the URL and transmits it to a remote server. Then, the server checks the URL against both the Conspiracy Resource dataset and the URLs dataset identified in Sec.~\ref{sec:urls_matching}. If the server finds a match, it generates an informative message, which the plug-in displays to the user.}

\reviewed{
The source code for both tools and additional technical implementation details are publicly available on GitHub~\cite{channel_checker_bot,conspiracy_alert_plugin}. Notably, we designed both tools to maintain user privacy by not retaining user information (\eg IP addresses) or details that could later associate a request with a user (\eg timestamps, Telegram client versions, or browsing history). The system only stores information about the analyzed channels.}



\section{Limitations}
\label{sec:limitation}
As mentioned in Sec.~\ref{sec:detection-methodology}, we build the \textit{Conspiracy Resource Dataset} gathering information from previous work focused on conspiracy theories. However, most of the scientific literature focuses on analyzing English content. This limitation could introduce bias, as conspiracy communities operating in other languages, such as Russian or Indian, might evade detection due to their use of non-English sources.
Moreover, other platforms not considered in our study (\eg Parler~\cite{bar2023finding}) are known to host conspiracy-related content. Unfortunately, we could not find works providing conspiracy-labeled datasets suitable for our study.

\reviewed{This study primarily utilizes link analysis to uncover monetization techniques and estimate the amount of money raised. This approach does not provide insight into monetization campaigns conducted through media content. For example, we observed various images shared on the channels promoting self-published books, paid events (\eg conferences), or training courses conducted by self-proclaimed gurus.
In Sec.~\ref{sec:crowd}, we analyzed campaigns that collected more than \$200,000 and found several that, despite being endorsed by conspiracy channels, do not show evidence of being connected to conspiracy theories. This suggests that other similar campaigns may exist among those we considered in this study.
Understanding why these campaigns are endorsed by conspiracy channels raises intriguing questions: Are they shared only to support the conspiracy narratives? Are their topics intentionally unrelated to conspiracy theories to appeal to a broader audience? Are administrators paid to promote these campaigns?
}

Throughout our investigation, we do not attempt to infer the direct link between the channel's administrator and the ultimate recipient of funds. In certain situations, this connection is clear, such as when the channel's name matches a profile on an external platform or in the case of affiliate program campaigns. However, in other instances, such as crowdfunding campaigns, it proves challenging to discern the ultimate objective of the channel's administrator. However, it is clear that someone is profiting and that the channels have a key role in fueling the conspiracy theories' money machine.

\section{Ethical considerations}
The dataset we analyze does not contain personal information like phone numbers or any media that could include adult content or copyrighted material. Furthermore, the channels mentioned in our study are publicly accessible and represent widely recognized public figures or entities. In our data collection process, we scraped web pages of the analyzed platforms. We adopt a careful approach to prevent flooding and ensure a minimal impact on these services by limiting the volume of requests submitted.

\section{Conclusion and future work}
In this work, we focused on understanding and quantifying how conspiracy theories raise funds by exploiting Telegram.
We started by identifying the conspiracy theory-related channels, analyzing a novel dataset we built by collecting previously validated resources from an extensive literature review that we publicly release~\cite{anonymous_groundtruth}.
This study revealed the alarming finding that almost 15\% of all Telegram channels in the TGDataset (17,829 channels) are linked to conspiracy theories.
Then, we discover that conspiracy theory-related channels actively seek to profit from their subscribers.
We categorize all the diverse monetization strategies we find in our dataset and dive into the analysis of the three most common.
Our study shows that conspiracy theories raised funds for about \$71 million by arranging crowdfunding campaigns.

As a future work, we believe it is interesting to conduct a more comprehensive analysis of the monetization strategies reported in Sec.~\ref{sec:other_strategies} to get deeper insights into the impact of monetization. 
Finally, another possible direction is analyzing the channels that use the same affiliate program ID  and those that share identical funding projects. This study could highlight the collaborative patterns presented by these channels and enable the identification of more fine-grained sub-communities.

\section*{Acknowledgments}
This work has been partially funded by projects: MUR National Recovery and Resilience Plan, SERICS (PE00000014); and ST3P (B83C24003210001) under the "Young Researchers 2024-SoE" Program funded by the Italian Ministry of University and Research (MUR). We express our gratitude to Lucia Fores for her contributions to the Channel Checker Bot and the Conspiracy Alert browser plugin.


\bibliographystyle{plain}

\bibliography{biblio}

\appendix 
\input{appendix}

\end{document}

%% file: appendix.tex
\section{ChatGPT prompt}
\begin{table}[H]
   \centering
   \small
   \caption{%
    Prompt used to ask GPT-4o whether the author of a
message containing a monetization link is requesting funds or
promoting a campaign or if they are discrediting a donation
or crowdfunding project.
  }\label{tab:chatgpt_prompt}
   \begin{tabular}{p{0.95\linewidth} }
      \toprule
    \textbf{Prompt}\\
      \midrule
The input provided is a text containing one or more URLs pointing to donation or crowdfunding platform.
Classify the message with 1 if it is asking directly or indirectly for donations, or to support a project.
Otherwise, return 0 if the message is discrediting or asking to report the recipient of the donation or the project.\\
    \bottomrule
    \end{tabular}
\end{table}

\section{Amazon content}

\begin{table}[H]
   \centering
   \small
   \caption{%
    The five types of products most frequently advertised by conspiracy channels on Amazon.
  }\label{tab:amazon_products}
   \begin{tabular}{l l l }
      \toprule
    \textbf{Category} & \textbf{URLs} & \textbf{Distinct products}\\
      \midrule
    Books \& eBooks & 27,523 (54.72\%) & 5,213 (30.37\%) \\
    Fashion \& Personal Care & 7,820 (15.55\%) & 3,700 (21.56\%) \\
    Home \& Living & 6,715 (13.35\%) & 3,688 (21.49\%) \\
    Electronics \& Technology & 4,189 (8.33\%) & 2,358 (13.74\%)\\
    Sports \& Outdoors & 1,618 (3.22\%) &  636 (3.71\%) \\

    \bottomrule
    \end{tabular}
\end{table}

\section{Platforms usage over time}

\begin{figure}[H]
    \centerline{\includegraphics[width=0.44\textwidth]{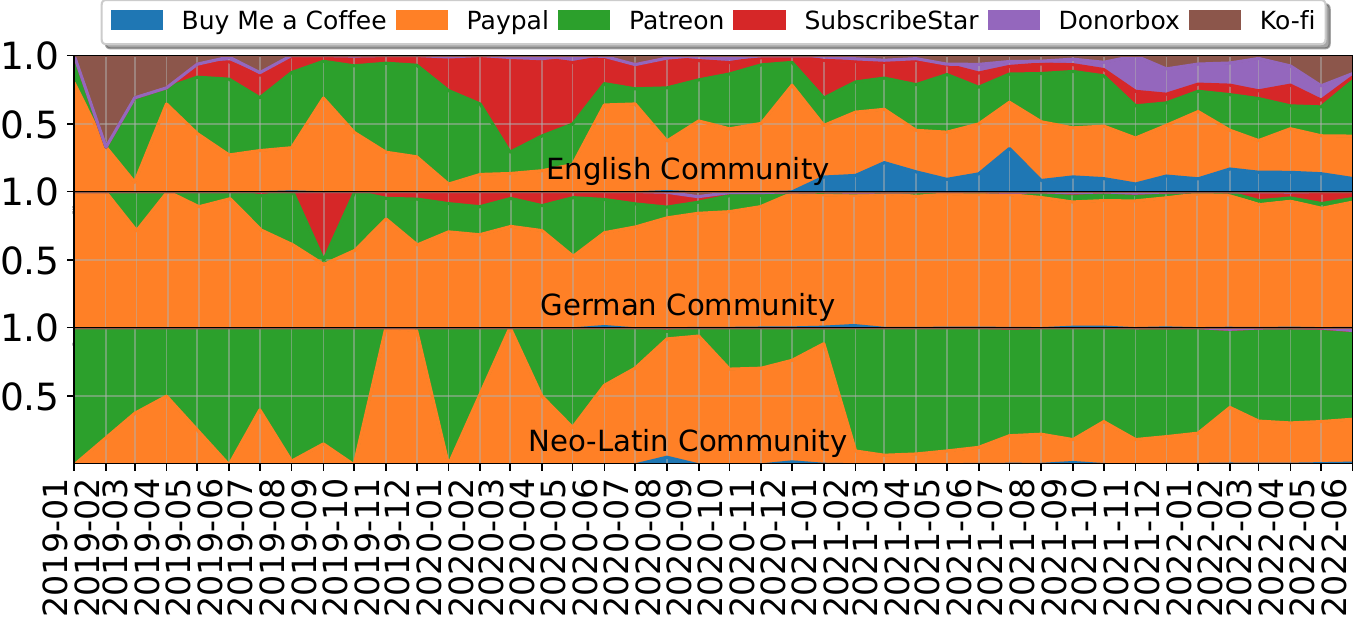}}
    \caption{Donation platforms usage over time.}
  \label{fig:platform_usage_over_time_donation}
\end{figure}

\begin{figure}[H]
    \centerline{\includegraphics[width=0.44\textwidth]{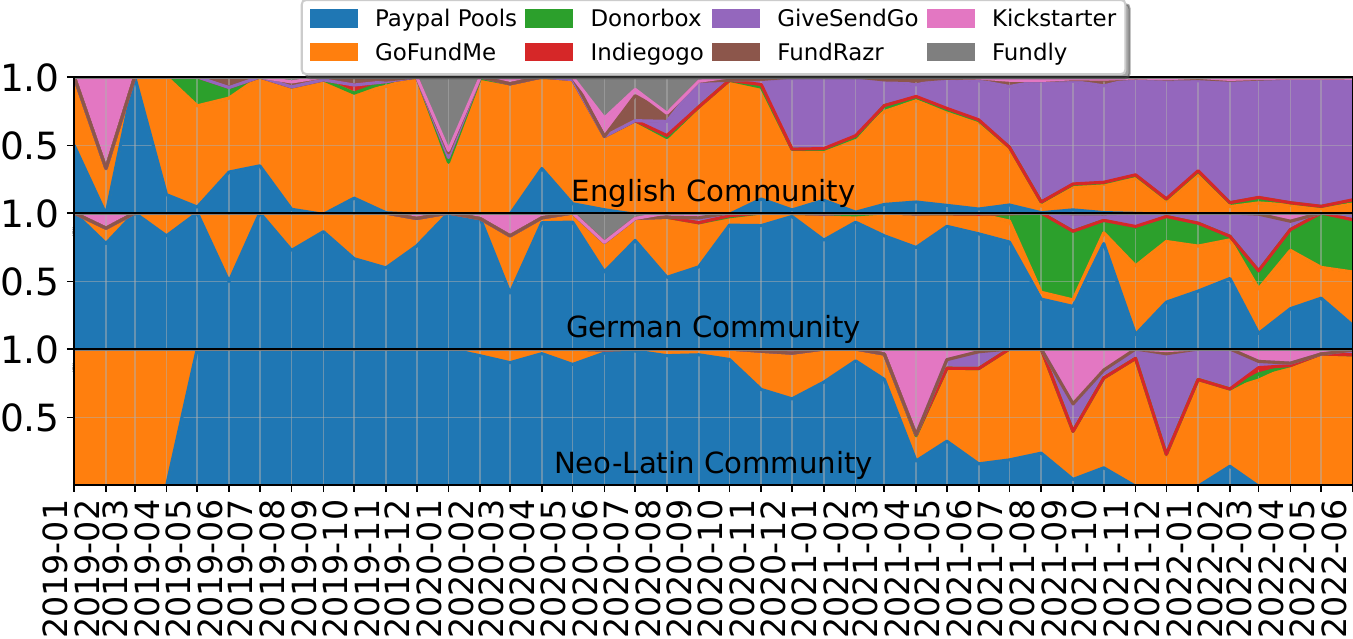}}
    \caption{Crowdfunding platforms usage over time.}
  \label{fig:platform_usage_over_time_crowdfunding}
\end{figure}

      


%% file: main.bbl
\begin{thebibliography}{10}

\bibitem{FakeNewsCorpus}
Fake news corpus.
\newblock \url{https://github.com/several27/FakeNewsCorpus}, 2023.

\bibitem{QAnonanonymous}
Qanon anonymous.
\newblock \url{www.patreon.com/qanonanonymous}, 2023.

\bibitem{TelegramUsers}
Telegram faq.
\newblock \url{https://telegram.org/faq}, 2023.

\bibitem{telegramconspiracy2}
France 24.
\newblock Germany weighs ban on telegram, tool of conspiracy theorists.
\newblock \url{https://f24.my/8KcH}, 2022.

\bibitem{fallodthecabal}
ADL.
\newblock Fall of the cabal.
\newblock \url{www.adl.org/glossary/fall-cabal}, 2023.

\bibitem{ahmed2020covid}
Wasim Ahmed, Josep Vidal-Alaball, Joseph Downing, Francesc~L{\'o}pez Segu{\'\i}, et~al.
\newblock Covid-19 and the 5g conspiracy theory: social network analysis of twitter data.
\newblock {\em Journal of medical internet research}, 2020.

\bibitem{aliapoulios2021large}
Max Aliapoulios, Emmi Bevensee, Jeremy Blackburn, Barry Bradlyn, Emiliano De~Cristofaro, Gianluca Stringhini, and Savvas Zannettou.
\newblock A large open dataset from the parler social network.
\newblock In {\em AAAI ICWSM}, 2021.

\bibitem{amazonaffiliate}
Amazon.
\newblock Associates program commission income statement.
\newblock \url{https://affiliate-program.amazon.com/help/node/topic/GRXPHT8U84RAYDXZ}, 2023.

\bibitem{amazonbooks}
Amazon.
\newblock Content guidelines for books.
\newblock \url{www.amazon.com/gp/help/customer/display.html?nodeId=201995150}, 2023.

\bibitem{amazoninfluencer}
Amazon.
\newblock Monetize your content with the amazon influencer program.
\newblock \url{https://affiliate-program.amazon.com/influencers}, 2023.

\bibitem{helpkids}
Inc American Education~Defenders.
\newblock Help protect our kids against the raw sewage of crt and other indoctrinations.
\newblock \url{www.fundrazr.com/ourfuture}, 2021.

\bibitem{ballard2022conspiracy}
Cameron Ballard, Ian Goldstein, Pulak Mehta, Genesis Smothers, Kejsi Take, Victoria Zhong, Rachel Greenstadt, Tobias Lauinger, and Damon McCoy.
\newblock Conspiracy brokers: understanding the monetization of youtube conspiracy theories.
\newblock In {\em ACM Web Conference}, 2022.

\bibitem{bar2023finding}
Dominik B{\"a}r, Nicolas Pr{\"o}llochs, and Stefan Feuerriegel.
\newblock Finding qs: Profiling qanon supporters on parler.
\newblock In {\em AAAI ICWSM}, 2023.

\bibitem{barker2020germany}
Tyson Barker.
\newblock Germany is losing the fight against qanon.
\newblock {\em Foreign Policy}, 2020.

\bibitem{twitterbanqanon}
BBC.
\newblock Twitter suspends 70,000 accounts linked to qanon.
\newblock \url{https://bbc.com/news/technology-55638558}, 2021.

\bibitem{gofundmeconvoy}
BBC.
\newblock Freedom convoy: Gofundme seizes funds of canada 'occupation'.
\newblock \url{https://bbc.com/news/world-us-canada-60267840}, 2022.

\bibitem{telegramconspiracy}
BBC.
\newblock The light: Inside the uk’s conspiracy theory newspaper that shares violence and hate.
\newblock \url{www.bbc.com/news/uk-65821747}, 2023.

\bibitem{bernstein20114chan}
Michael Bernstein, Andr{\'e}s Monroy-Hern{\'a}ndez, Drew Harry, Paul Andr{\'e}, Katrina Panovich, and Greg Vargas.
\newblock 4chan and /b: An analysis of anonymity and ephemerality in a large online community.
\newblock In {\em AAAI ICWSM}, 2011.

\bibitem{bleakley2023panic}
Paul Bleakley.
\newblock Panic, pizza and mainstreaming the alt-right: A social media analysis of pizzagate and the rise of the qanon conspiracy.
\newblock {\em Current sociology}, 2023.

\bibitem{blekanov2021detection}
Ivan Blekanov, Svetlana~S Bodrunova, and Askar Akhmetov.
\newblock Detection of hidden communities in twitter discussions of varying volumes.
\newblock {\em Future Internet}, 2021.

\bibitem{bond2023rise}
Bayleigh~Elaine Bond and Ryan Neville-Shepard.
\newblock The rise of presidential eschatology: Conspiracy theories, religion, and the january 6th insurrection.
\newblock {\em American Behavioral Scientist}, 2023.

\bibitem{borodin2001finding}
Allan Borodin, Gareth~O Roberts, Jeffrey~S Rosenthal, and Panayiotis Tsaparas.
\newblock Finding authorities and hubs from link structures on the world wide web.
\newblock In {\em ACM Web Conference}, 2001.

\bibitem{broniatowski2023measuring}
David~A Broniatowski, Kevin~T Greene, Nilima Pisharody, Daniel~J Rogers, and Jacob~N Shapiro.
\newblock Measuring the monetization strategies of websites with application to pro-and anti-vaccine communities.
\newblock {\em Scientific reports}, 2023.

\bibitem{pyramidpower}
Derek Broze.
\newblock Help us finish the pyramid of power documentary series!
\newblock \url{https://fnd.us/pyramidofpowerdoc}, 2021.

\bibitem{chachra2015affiliate}
Neha Chachra, Stefan Savage, and Geoffrey~M Voelker.
\newblock Affiliate crookies: Characterizing affiliate marketing abuse.
\newblock In {\em ACM IMC}, 2015.

\bibitem{cheng2014understanding}
Xu~Cheng, Fatourechi Mehrdad, Xiaoqiang Ma, Cong Zhang, and Jiangchuan Liu.
\newblock Understanding the youtube partners and their data: Measurement and analysis.
\newblock {\em China Communications}, 2014.

\bibitem{coronadoc1}
Robert Cibis.
\newblock Corona.film.
\newblock \url{www.indiegogo.com/projects/corona-film?create\_edit=true}, 2020.

\bibitem{cinelli2022conspiracy}
Matteo Cinelli, Gabriele Etta, and Michele Avalle~et. al.
\newblock Conspiracy theories and social media platforms.
\newblock {\em Current Opinion in Psychology}, 2022.

\bibitem{clarkunderstanding}
Sam Clark and Anna Zaitsev.
\newblock Understanding youtube communities via subscription-based channel embeddings.
\newblock {\em arXiv preprint}, 2020.

\bibitem{clarke2019conspiracy}
Steve Clarke.
\newblock Conspiracy theories and conspiracy theorizing.
\newblock In {\em Conspiracy Theories}. Routledge, 2019.

\bibitem{cleland2020charismatic}
Jamie Cleland.
\newblock Charismatic leadership in a far-right movement: an analysis of an english defence league message board following the resignation of tommy robinson.
\newblock {\em Social Identities}, 2020.

\bibitem{flatearth}
CNN.
\newblock The flat-earth conspiracy is spreading around the globe. does it hide a darker core?
\newblock \url{www.cnn.com/2019/11/16/us/flat-earth-conference-conspiracy-theories-scli-intl}, 2019.

\bibitem{langdetect}
Michal~Mimino Danilak.
\newblock langdetect.
\newblock \url{https://pypi.org/project/langdetect/}, 2023.

\bibitem{de2020tracing}
Daniel De~Zeeuw, Sal Hagen, Stijn Peeters, and Emilija Jokubauskaite.
\newblock Tracing normiefication: A cross-platform analysis of the qanon conspiracy theory.
\newblock {\em First Monday}, 2020.

\bibitem{demuru2020conspiracy}
Paolo Demuru.
\newblock Conspiracy theories, messianic populism and everyday social media use in contemporary brazil a glocal semiotic perspective.
\newblock {\em Glocalism}, 2020.

\bibitem{douglas2017psychology}
Karen~M Douglas, Robbie~M Sutton, and Aleksandra Cichocka.
\newblock The psychology of conspiracy theories.
\newblock {\em Current directions in psychological science}, 2017.

\bibitem{ebayaffiliate}
eBay.
\newblock Global rate card.
\newblock \url{https://partnernetwork.ebay.com/our-program/rate-card}, 2023.

\bibitem{engel2022characterizing}
Kristen Engel, Yiqing Hua, Taixiang Zeng, and Mor Naaman.
\newblock Characterizing reddit participation of users who engage in the qanon conspiracy theories.
\newblock {\em ACM HCI}, 2022.

\bibitem{moonlandinghoax}
Emily Fales, Lauryn Lintner, Mason Runkel, and Paola Ariza.
\newblock The moon landing hoax.
\newblock 2020.

\bibitem{fernbach2023conspiracy}
Philip~M Fernbach and Jonathan~E Bogard.
\newblock Conspiracy theory as individual and group behavior: Observations from the flat earth international conference.
\newblock {\em Topics in Cognitive Science}, 2023.

\bibitem{garry2021qanon}
Amanda Garry, Samantha Walther, Rukaya Rukaya, and Ayan Mohammed.
\newblock Qanon conspiracy theory: examining its evolution and mechanisms of radicalization.
\newblock {\em Journal for Deradicalization}, 2021.

\bibitem{freedomgofundme}
GoFundMe.
\newblock Gofundme to refund all freedom convoy 2022 donations.
\newblock \url{www.gofundme.com/f/taking-back-our-freedom-convoy-2022}, 2022.

\bibitem{gordon2022freedom}
Todd Gordon.
\newblock The freedom convoy, the resurgence of the far right, and the crisis of the petty bourgeoisie.
\newblock {\em Studies in Political Economy}, 2022.

\bibitem{sabmykguardian}
The Guardian.
\newblock Unmasked: man behind cult set to replace qanon.
\newblock \url{www.theguardian.com/us-news/2021/mar/20/revealed-man-behind-fast-growing-cult-becoming-the-new-qanon-sabmyk-network}, 2021.

\bibitem{hanley2022no}
Hans~WA Hanley, Deepak Kumar, and Zakir Durumeric.
\newblock No calm in the storm: investigating qanon website relationships.
\newblock In {\em AAAI ICWSM}, 2022.

\bibitem{hoseini2023globalization}
Mohamad Hoseini, Philipe Melo, Fabricio Benevenuto, Anja Feldmann, and Savvas Zannettou.
\newblock On the globalization of the qanon conspiracy theory through telegram.
\newblock In {\em ACM WEBSCI}, 2023.

\bibitem{imhoff2020bioweapon}
Roland Imhoff and Pia Lamberty.
\newblock A bioweapon or a hoax? the link between distinct conspiracy beliefs about the coronavirus disease (covid-19) outbreak and pandemic behavior.
\newblock {\em Social Psychological and Personality Science}, 2020.

\bibitem{scamcampaign2}
Indiegogo.
\newblock Antivirbag-portable air cleaner, ionizer, ozonizer.
\newblock \url{https://igg.me/at/antivirbag}, 2020.

\bibitem{capitolhill3}
April Jensen.
\newblock God bless america, free my j6er.
\newblock \url{www.givesendgo.com/G26FY}, 2021.

\bibitem{capitolhill1}
Sommer~B Johnson.
\newblock Stand with paul.
\newblock \url{www.fundly.com/stand-4-paul}, 2021.

\bibitem{kleinberg1999hubs}
Jon~M Kleinberg.
\newblock Hubs, authorities, and communities.
\newblock {\em ACM CSUR}, 1999.

\bibitem{lamorgia2023trap}
Massimo La~Morgia, Alessandro Mei, and Alberto~Maria Mongardini.
\newblock It's a trap! detection and analysis of fake channels on telegram.
\newblock In {\em IEEE ICWS}, 2023.

\bibitem{la2023tgdataset}
Massimo La~Morgia, Alessandro Mei, and Alberto~Maria Mongardini.
\newblock Tgdataset: Collecting and exploring the largest telegram channels dataset.
\newblock In {\em ACM SIGKDD}, 2025.

\bibitem{la2018pretending}
Massimo La~Morgia, Alessandro Mei, Alberto~Maria Mongardini, and Jie Wu.
\newblock Pretending to be a vip! characterization and detection of fake and clone channels on telegram.
\newblock {\em ACM TWEB}, 2024.

\bibitem{la2023doge}
Massimo La~Morgia, Alessandro Mei, Francesco Sassi, and Julinda Stefa.
\newblock The doge of wall street: Analysis and detection of pump and dump cryptocurrency manipulations.
\newblock {\em ACM TOIT}, 2023.

\bibitem{ledwich2020algorithmic}
Mark Ledwich and Anna Zaitsev.
\newblock Algorithmic extremism: Examining youtube's rabbit hole of radicalization.
\newblock {\em First Monday}, 2020.

\bibitem{littrell2024political}
Shane Littrell, Casey Klofstad, and Joseph~E Uscinski.
\newblock The political, psychological, and social correlates of cryptocurrency ownership.
\newblock {\em PloS one}, 2024.

\bibitem{scamcampaign1}
Team McAfee.
\newblock Help team mcafee give back to savethechildren venezuela thanksgiving feast4food.
\newblock \url{www.fundrazr.com/Team\_McAfee}, 2021.

\bibitem{meder2021online}
Theo Meder et~al.
\newblock Online coping with the first wave: Covid humor and rumor on dutch social media (march--july 2020).
\newblock {\em Electronic Journal of Folklore}, 2021.

\bibitem{mekacher2022can}
Amin Mekacher and Antonis Papasavva.
\newblock "i can’t keep it up." a dataset from the defunct voat. co news aggregator.
\newblock In {\em AAAI ICWSM}, 2022.

\bibitem{mohammed2019conspiracy}
Shaheed~N Mohammed.
\newblock Conspiracy theories and flat-earth videos on youtube.
\newblock {\em The Journal of Social Media in Society}, 2019.

\bibitem{muhammad2021study}
Mohd Razman~Achmadi Muhammad and Noor Nirwandy.
\newblock A study on donald trump twitter remark: a case study on the attack of capitol hill.
\newblock {\em Journal of Media and Information Warfare (JMIW)}, 2021.

\bibitem{zerohedgeban}
CBS News.
\newblock Twitter bans zero hedge after it posts coronavirus conspiracy theory.
\newblock \url{www.cbsnews.com/news/twitter-bans-zero-hedge-coronavirus-conspiracy-theory/}, 2020.

\bibitem{nizzoli2020charting}
Leonardo Nizzoli, Serena Tardelli, Marco Avvenuti, Stefano Cresci, Maurizio Tesconi, and Emilio Ferrara.
\newblock Charting the landscape of online cryptocurrency manipulation.
\newblock {\em IEEE Access}, 2020.

\bibitem{aliapoulios2021gospel}
Antonis Papasavva, Max Aliapoulios, Cameron Ballard, Emiliano De~Cristofaro, Gianluca Stringhini, Savvas Zannettou, and Jeremy Blackburn.
\newblock The gospel according to q: Understanding the qanon conspiracy from the perspective of canonical information.
\newblock In {\em AAAI ICWSM}, 2021.

\bibitem{papasavva2021qoincidence}
Antonis Papasavva, Jeremy Blackburn, Gianluca Stringhini, Savvas Zannettou, and Emiliano~De Cristofaro.
\newblock “is it a qoincidence?”: An exploratory study of qanon on voat.
\newblock In {\em ACM Web Conference}, 2021.

\bibitem{papasavva2020raiders}
Antonis Papasavva, Savvas Zannettou, Emiliano De~Cristofaro, Gianluca Stringhini, and Jeremy Blackburn.
\newblock Raiders of the lost kek: 3.5 years of augmented 4chan posts from the politically incorrect board.
\newblock In {\em AAAI ICWSM}, 2020.

\bibitem{paypaldonate}
Paypal.
\newblock Donate button.
\newblock \url{www.paypal.com/donate/buttons}, 2023.

\bibitem{phadke2021makes}
Shruti Phadke, Mattia Samory, and Tanushree Mitra.
\newblock What makes people join conspiracy communities? role of social factors in conspiracy engagement.
\newblock {\em ACM HCI}, 2021.

\bibitem{phadke2022pathways}
Shruti Phadke, Mattia Samory, and Tanushree Mitra.
\newblock Pathways through conspiracy: the evolution of conspiracy radicalization through engagement in online conspiracy discussions.
\newblock In {\em AAAI ICWSM}, 2022.

\bibitem{capitolhill2}
Jodi Reffitt.
\newblock Reffitt family fund.
\newblock \url{www.givesendgo.com/G23DE}, 2021.

\bibitem{us_far_right_qanon}
Kevin Roose.
\newblock What is qanon, the viral pro-trump conspiracy theory?
\newblock \url{www.nytimes.com/article/what-is-qanon.html}, 2021.

\bibitem{disclosetv}
Elizabeth Schumacher.
\newblock Disclose.tv: English disinformation made in germany.
\newblock \url{www.dw.com/en/disclosetv-english-disinformation-made-in-germany/a-60694332}, 2023.

\bibitem{selenium}
Selenium.
\newblock Selenium automates browsers. that's it!
\newblock \url{www.selenium.dev}, 2023.

\bibitem{shehabat2017encrypted}
Ahmad Shehabat, Teodor Mitew, and Yahia Alzoubi.
\newblock Encrypted jihad: Investigating the role of telegram app in lone wolf attacks in the west.
\newblock {\em JSS}, 2017.

\bibitem{sliwa2024case}
Karolina Sliwa, Ema Kusen, and Mark Strembeck.
\newblock A case study comparing twitter communities detected by the louvain and leiden algorithms during the 2022 war in ukraine.
\newblock In {\em ACM Web Conference}, 2024.

\bibitem{stempel2007media}
Carl Stempel, Thomas Hargrove, and Guido~H Stempel~III.
\newblock Media use, social structure, and belief in 9/11 conspiracy theories.
\newblock {\em Journalism \& Mass Communication Quarterly}, 2007.

\bibitem{coronadoc2}
Maria Susana.
\newblock Crimes against humanity.
\newblock \url{www.indiegogo.com/projects/crimes-against-humanity?create\_edit=true}, 2020.

\bibitem{channel_checker_bot}
SystemsLab-Sapienza.
\newblock Channel checker bot.
\newblock \url{https://github.com/SystemsLab-Sapienza/channel-checker-bot}, 2025.

\bibitem{anonymous_groundtruth}
SystemsLab-Sapienza.
\newblock Conspiracy resource and url datasets.
\newblock \url{https://github.com/SystemsLab-Sapienza/conspiracy-dataset-telegram}, 2025.

\bibitem{conspiracy_alert_plugin}
SystemsLab-Sapienza.
\newblock Conspiracyalert plugin.
\newblock \url{https://github.com/SystemsLab-Sapienza/conspiracy-alert-plugin}, 2025.

\bibitem{telegramads}
Telegram.
\newblock Telegram ad platform.
\newblock \url{https://promote.telegram.org}, 2023.

\bibitem{traag2019louvain}
Vincent~A Traag, Ludo Waltman, and Nees~Jan Van~Eck.
\newblock From louvain to leiden: guaranteeing well-connected communities.
\newblock {\em Scientific reports}, 2019.

\bibitem{trujillo2020bitchute}
Milo Trujillo, Maur{\'\i}cio Gruppi, Cody Buntain, and Benjamin~D Horne.
\newblock What is bitchute? characterizing the.
\newblock In {\em ACM Hypertext}, 2020.

\bibitem{tuters2018post}
Marc Tuters, Emilija Jokubauskait{\.e}, and Daniel Bach.
\newblock Post-truth protest: How 4chan cooked up the pizzagate bullshit.
\newblock {\em M/c Journal}, 2018.

\bibitem{van2017conspiracy}
Jan-Willem Van~Prooijen and Karen~M Douglas.
\newblock Conspiracy theories as part of history: The role of societal crisis situations.
\newblock {\em Memory studies}, 2017.

\bibitem{qanonvoat}
The Verge.
\newblock Reddit has banned the qanon conspiracy subreddit r/greatawakening.
\newblock \url{www.theverge.com/2018/9/12/17851938/reddit-qanon-ban-conspiracy-subreddit-greatawakening}, 2023.

\bibitem{oliverjanich}
VICE.
\newblock Germany’s ‘biggest qanon mouthpiece’ arrested in the philippine.
\newblock \url{www.vice.com/en/article/n7zexk/oliver-janich-germany-philippines}, 2022.

\bibitem{visser2024crowdfunding}
Rebecca Visser.
\newblock Crowdfunding conspiracists: Grassroots giving to january 6 participants.
\newblock 2024.

\bibitem{telegramconspiracy1}
VSQUARE.
\newblock Telegram, the free zone for disinformation and conspiracies.
\newblock \url{https://vsquare.org/telegram-the-free-zone-for-disinformation-and-conspiracies/}, 2023.

\bibitem{freedomconvoy}
W.
\newblock Freedomconvoy2022.
\newblock \url{www.givesendgo.com/FreedomConvoy2022}, 2021.

\bibitem{coronadoc3}
W.
\newblock The big reset.
\newblock \url{www.kickstarter.com/projects/thebigreset/the-big-reset}, 2021.

\bibitem{wagnsson2023paperboys}
Charlotte Wagnsson.
\newblock The paperboys of russian messaging: Rt/sputnik audiences as vehicles for malign information influence.
\newblock {\em Information, communication \& society}, 2023.

\bibitem{weerasinghe2020pod}
Janith Weerasinghe, Bailey Flanigan, Aviel Stein, Damon McCoy, and Rachel Greenstadt.
\newblock The pod people: Understanding manipulation of social media popularity via reciprocity abuse.
\newblock In {\em ACM Web Conference}, 2020.

\bibitem{willman1998traversing}
Skip Willman.
\newblock Traversing the fantasies of the jfk assassination: Conspiracy and contingency in don delillo's" libra".
\newblock {\em Contemporary Literature}, 1998.

\bibitem{yayla2017telegram}
Ahmet~S Yayla and Anne Speckhard.
\newblock Telegram: The mighty application that isis loves.
\newblock {\em International Center for the Study of Violent Extremism}, 2017.

\bibitem{zeng2021conceptualizing}
Jing Zeng and Mike~S Sch{\"a}fer.
\newblock Conceptualizing “dark platforms”. covid-19-related conspiracy theories on 8kun and gab.
\newblock {\em Digital Journalism}, 2021.

\end{thebibliography}
